\documentclass[12pt]{iopart}

\usepackage{iopams}
\usepackage{dcolumn}
\usepackage{color}
\usepackage[final]{graphicx}
\usepackage{bm}
\usepackage{comment}

\newcommand{\be}{\begin{equation}}
\newcommand{\ee}{\end{equation}}
\newcommand{\bea}{\begin{eqnarray}}
\newcommand{\eea}{\end{eqnarray}}
\newcommand{\bit}{\begin{itemize}}
\newcommand{\eit}{\end{itemize}}

\newcommand{\bfl}{\begin{flushright}}
\newcommand{\efl}{\end{flushright}}

\newcommand{\non}{\nonumber \\}

\newcommand{\ra}{\rangle}
\newcommand{\la}{\langle}

\newcommand{\ah}{\hat{a}}

\excludecomment{pdffigure}

\newcommand{\eqref}[1]{(\ref{#1})}

\begin{document}

\bibliographystyle{../../../../julianNormal}

\title[Atom interferometry with trapped Bose-Einstein condensates]%
{Atom interferometry with trapped Bose-Einstein condensates: Impact of atom--atom interactions}

\author{Julian Grond$^1$, Ulrich Hohenester$^1$, Igor Mazets$^{2,3,4}$, and J\"org Schmiedmayer$^2$ }
\address{$^1$ Institut f\"ur Physik, Karl--Franzens--Universit\"at Graz, Universit\"atsplatz 5, 8010 Graz, Austria}
\address{$^2$ Atominstitut, TU--Wien, Stadionallee 2, 1020 Vienna, Austria}
\address{$^3$ A.F. Ioffe Physico-Technical Institute, 194021 St. Petersburg, Russia}
\address{$^4$ Wolfgang Pauli Institut, Nordbergstrasse 15, 1090 Vienna, Austria}
\ead{julian.grond@uni-graz.at}

\date{\today}

\begin{abstract}

Interferometry with ultracold atoms promises the possibility of ultraprecise and ultrasensitive measurements in many ﬁelds of physics, and is
the basis of our most precise atomic clocks.   Key to a high sensitivity is the possibility to achieve long measurement times and precise readout. Ultra cold atoms can be precisely manipulated at the quantum level, held for very long times in traps, and would therefore be an ideal setting for interferometry. In this paper we discuss how the non-linearities from atom-atom interactions on one hand allow to efficiently produce squeezed states for enhanced readout, but on the other hand result in phase diffusion which limits the phase accumulation time. We find that low dimensional geometries are favorable, with two-dimensional (2D) settings giving the smallest contribution of phase diffusion caused by atom-atom interactions.   Even for time sequences generated by optimal control the achievable minimal detectable interaction energy $\Delta E^{\rm min}$ is on the order of $10^{-4} \mu$, where $\mu$ is the chemical potential of the BEC in the trap. From there we have to conclude that for more precise measurements with atom interferometers more sophisticated strategies, or turning off the interaction induced dephasing during the phase accumulation stage, will be necessary.

\end{abstract}



\submitto{\NJP}

\newpage
\tableofcontents
\newpage

\section{Introduction}

Interferometry is the method of choice in achieving the most precise measurements, or when trying to detect the most feeble effects or signals.  Interferometry with matter waves \cite{cronin:09} became a very versatile tool with many applications ranging from precision experiments to fundamental studies.

In an interferometer, an incoming \textit{'beam'} (light or matter ensemble) is split into two parts (pathways), which can be separated in either internal state space or real space. The splitting process prepares the two paths with a well-defined relative phase $\delta \theta(t=0) = \theta_1(t=0) - \theta_2(t=0)$.  After the splitting they  evolve separately, and can accumulate different phases $\theta_1(t)$ and $\theta_2(t)$ due to the different physical settings they evolve in. Finally, in the recombination process after time $T$ the relative phase $\delta \theta (T) = \theta_1(T) - \theta_2(T)$ accumulated in the two paths can be read out.

The sensitivity of an interferometer measurement depends now on two distinct points: How good can the phase difference $\delta \theta$ be measured, and how long can one accumulate a phase difference in the split paths.  For perfect read-out contrast and standard (binomial) splitting and recombination procedures, the uncertainty in determining $\Delta \theta$ is given by the standard quantum limit $\Delta \theta = 1/\sqrt{N}$, where $N$ is the number of registered counts (e.g. atom detections). The second point concerns the question of how long the beams can be kept in the 'interaction region' of the distinguishable interferometer arms, and for how long the paths stay coherent.  Ultra cold (degenerate) atoms can be held and manipulated in well controlled traps and guides, and therefore promise ultimate precision and sensitivity for interferometry.  Both dipole traps \cite{grimm:00} and atom chips \cite{folman:00,folman:02,Reichel:2002,fortagh:07} have been used to analyze different interferometer geometries \cite{hinds:01,haensel:01,andersson:02,Kreutzmann:04} and employed for experimental demonstration of splitting \cite{Houde:00,cassettari:00,cassettari:00b,Dumke:02} and interference  \cite{shin:04,schumm:05,wang:05,albiez:05,hofferberth:07,jo:07,jo2:07,hofferberth:08,esteve:08,bohi:09} with trapped or guided ultra cold atoms. 

The power of interferometry lies in the precision and robustness of the phase evolution, which provides the measurement stick. This robustness of the phase evolution is based on the linearity of time propagation in the different paths, which is the case for most light interferometers, atom interferometers \cite{cronin:09} with weak and dilute beams, or neutron interferometers \cite{rauch:74}. Measurements loose precision when this robustness of the phase evolution cannot be guaranteed.  This is the case if the phase evolution in the paths depends on the intensity (density), that is, when the time evolution becomes non-linear. Atom optics is fundamentally non-linear, the non-linearity being created by the interaction between atoms.  For ultra cold (degenerate) trapped Bose gases this mean field energy associated with the atom-atom interaction can even dominate the time evolution. Consequently, in many interferometer experiments with trapped atoms the atom-atom interaction creates a non-linearity in the time propagation, and the accumulated phase depends on the local atomic densities. Thus, number fluctuations induced by the splitting cause phase diffusion \cite{lewenstein:96,javanainen:97}, which currently limits the coherence and sensitivity of interferometers with trapped atoms much more then decoherence coming from other sources like the surface \cite{henkel:03,skagerstam:06,hohenester.pra2:07}. 

In the present manuscript we will discuss the physics that leads to degradation of performance of an atom interferometer with trapped atoms, and how one can counteract it by using optimal input states.  We first discuss the performance of a trapped atom interferometer in the simplest two-mode model. This will allow us to illustrate the basic physics.  We then investigate how this simple two-mode model has to be modified when taking into account the many-body structure of the wave function.  Optimal control techniques are applied to prepare the desired input states \cite{hohenester.pra:07,grond.pra:09,grond.pra:09b}. 

In our calculations we always assume zero temperature. The effects of additional dephasing and decoherence due to fundamental quantum noise and due to thermal excitations at finite temperature will be discussed in the last section.

\section{Two-mode model description of atom interferometry}\label{sec:2mode}

When atom interferometry is performed with a trapped Bose Einstein condensate (BEC), we consider the following key stages: The splitting stage (with the time duration $T_{\rm split}$), where the condensate wave function is split into two parts, the phase accumulation stage ($T_{\rm phase}$), where the atoms in one arm of the interferometer experience an interaction with some weak (classical) field, and finally the read-out stage ($T_{\rm tof}$), where the phase accumulated is measured after the condensates have expanded in time-of-flight (TOF). 

In our theoretical description, we start by introducing a simple but generic description scheme of an interferometer in terms of a two-mode model for the split condensate. Such a model has been also proven successful for the description of interference with spin squeezed states \cite{oblak:05,fernholz:08} and of condensates in double wells \cite{albiez:05,esteve:08}. To properly account for the many-boson wave function, we introduce the field operator in second-quantized form \cite{leggett:01}
\begin{equation}\label{eq:fieldop}
  \hat\Psi(x)=\hat a_L\,\phi_L(x)+\hat a_R\,\phi_R(x)\,.
\end{equation}
Here, $a_{L,R}^\dagger$ are bosonic field operators that create an atom in the left or right well, with wavefunctions $\phi_{L,R}(x)$, respectively. In many cases, we can properly describe the dynamics of the interacting many-boson system by means of a generic \textit{two-mode hamiltonian} \footnote{From now on we use conveniently scaled time, length, and energy units \cite{grond.pra:09b}, unless stated differently. First we set $\hbar=1$ and scale the energy and time according to a harmonic oscillator with confinement length $a_{ho} = \sqrt{\hbar/(m\omega_{ho})}=1$ $\mu m$ and energy $\hbar \omega_{ho}$. Our considerations deal with Rb atoms, we therefore measure mass in units of the $^{87}$Rb atom, and then time is measured in units of $1/\omega_{ho}=1.37$ ms and energy in units of $\hbar \omega_{ho}=2\pi\cdot 116.26$ Hz.}  \cite{milburn:97,javanainen:99,grond.pra:09b}
\begin{equation}\label{eq:hamtwomode}
  \hat H=-\frac{\Omega}2\left(\hat a_L^\dagger \hat a_R^{\phantom\dagger}+
  \hat a_R^\dagger \hat a_L^{\phantom\dagger}\right)+
  \kappa\left(
  \hat a_L^\dagger\hat a_L^\dagger\hat a_L^{\phantom\dagger}\hat a_L^{\phantom\dagger}+
  \hat a_R^\dagger\hat a_R^\dagger\hat a_R^{\phantom\dagger}\hat a_R^{\phantom\dagger}
  \right)\,.
\end{equation}
Here $\Omega$ describes a tunnelling process, that allows the atoms to hop between the two wells and $\kappa$ is the non-linear atom-atom interaction, that energetically penalizes states with a high atom-number imbalance between the left and right well. We treat $\Omega$ as a free parameter, but choose a fixed $\kappa\approx U_0/2$. $U_0$ is the effective one-dimensioanl (1D) interaction strength along the direction where the potential is split into a double well.
A more accurate way of how to relate $\kappa$ and $U_0$ will be given  in section~\ref{sec:MCTDHB}.

In the example we discuss in this manuscript the initial state of the interferometry sequence is prepared by deforming a confinement potential from a single to a double well. This corresponds to a situation where one starts with the two-mode hamiltonian of equation~\eqref{eq:hamtwomode} for a large tunnel coupling, $\Omega\gg\kappa$, and then turns off $\Omega$, as a consequence of the reduced spatial overlap of the wavefunctions $\phi_{L,R}(x)$. 

In the beginning of the splitting sequence tunnelling dominates over the non-linear interaction, and all atoms reside in the bonding orbital $\phi_L(x)+\phi_R(x)$. This results in a binomial atom number distribution. When $\Omega$ is turned off sufficiently fast, and the dynamics due to the non-linear coupling plays no significant role, the orbitals $\phi_{L,R}(x)$ become spatially separated, but the atom number distribution remains binomial. In contrast, when $\Omega$ is turned off sufficiently slowly, the system can adiabatically follow the groundstate of the Hamiltonian \eqref{eq:hamtwomode} and ends up approximately in a Fock state. As we will discuss below, such states with reduced atom number fluctuations are appealing for the purpose of atom interferometry.

\subsection{Pseudospin operators and Bloch sphere}

A convenient representation of the two-mode model for a many-boson system is in terms of angular-momentum operators \cite{milburn:97,javanainen:99,hohenester.fdp:09}. Quite generally, the internal state of an ensemble of atoms which are allowed to occupy two states (here left and right) can be described as a collective (pseudo)spin ${\hat{\bm J}} = \sum_{i=1}^N {\hat{\bm j}_i}$, which is the sum of the individual spins of all atoms. Here the total angular momentum is $N/2$, and the projection $m$ on the $z$-axis corresponds to states where, starting from a state where the left and right well are each populated with $N/2$ atoms, $m$ atoms are promoted from the right to the left well. Through the Schwinger boson representation
\begin{equation}\label{eq:pseudospin}
  \mbox{\hspace*{-2cm}}
  \hat J_x=\frac 12\left(
    \hat a_L^\dagger \hat a_R^{\phantom\dagger}+
    \hat a_R^\dagger \hat a_L^{\phantom\dagger}\right)\,,\quad
  \hat J_y=-\frac i2\left(
    \hat a_L^\dagger \hat a_R^{\phantom\dagger}-
    \hat a_R^\dagger \hat a_L^{\phantom\dagger}\right)\,,\quad
  \hat J_z = \frac 12\left(
    \hat a_L^\dagger \hat a_L^{\phantom\dagger}-
    \hat a_R^\dagger \hat a_R^{\phantom\dagger}\right)
\end{equation}
we can establish a link between the field operators $\hat a_{L,R}$ and the pseudo-spin operators. $\hat J_x$ promotes an atom from the left to the right well, or vice versa, and $\hat J_z$ gives half the atom number difference between the two wells.

\begin{figure}
\center\includegraphics[width=0.5\columnwidth]{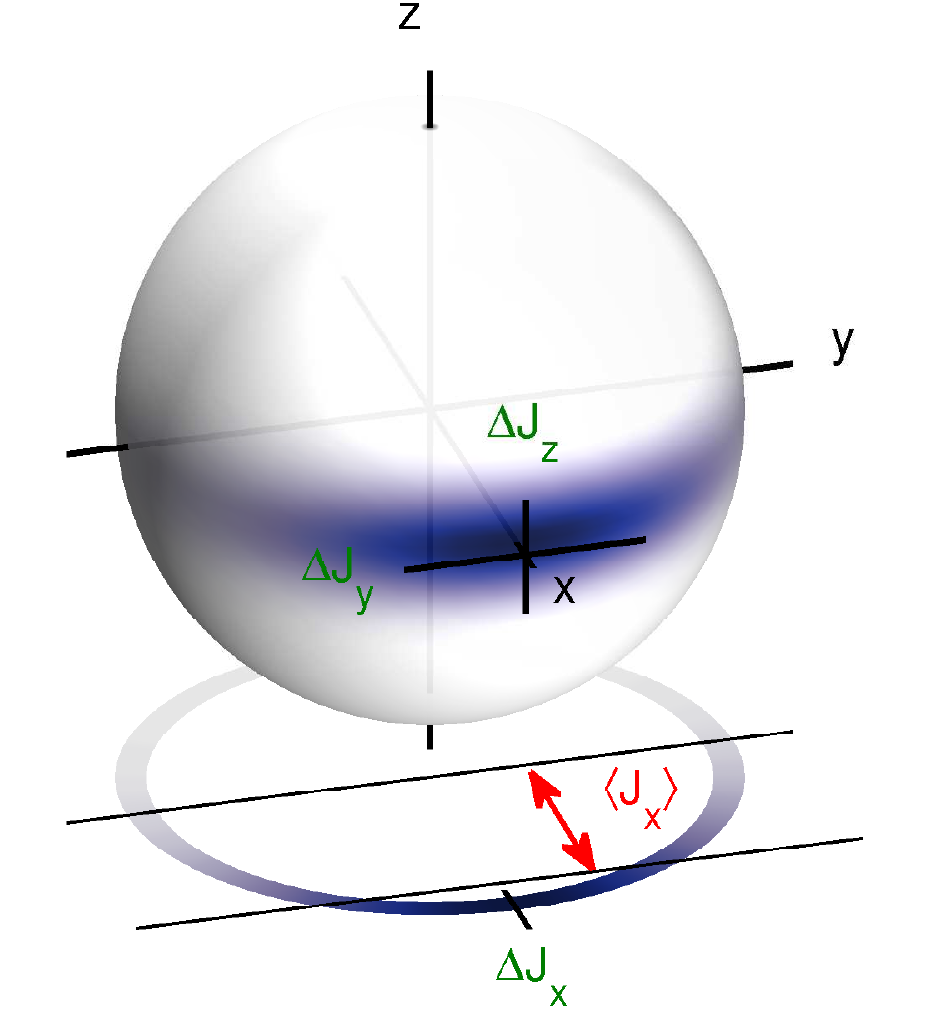}
\caption{Bloch sphere representation of a number squeezed state with squeezing factor (defined in text) $\xi_N\approx0.2$. $\Delta J_z$ corresponds to number squeezing, and $\Delta J_y$ is proportional to phase squeezing. In the ring below the sphere we show the polarization of the state along the x-axis, which is proportional to the coherence factor $\alpha=2\langle \hat J_x\rangle/N$. For squeezed states there is also a noise $\Delta J_x$ in the polarization of the state on the equator. }\label{fig:blochintro}
\end{figure}

One can map the two-mode wave function onto the Bloch sphere \cite{arecchi:72}, which provides an extremely useful visualization tool for the purpose of atom interferometry (see figure~\ref{fig:blochintro}).  A state on the north pole corresponds to all atoms residing in the left well and a state on the south pole to all atoms in the right well. All atoms in the bonding orbital corresponds to a state localized around $x=N/2$. This is a product state where the atoms are totally uncorrelated. In this state the quantum noise is evenly distributed among $ \Delta J_y  = \Delta J_z  = {\sqrt{N}}/{2}$, i.e., it has equal uncertainty in number difference (measured along the $z$-axis) and in the conjugate phase observable (measured around the equator of the sphere). Similar to optics with photons, or as discussed in the context of spin squeezing, quantum correlations can reduce the variance of one spin quadrature, for a given angle $\phi$, $\hat J_{\phi}=\cos\phi\hat J_z+\sin\phi\hat J_y$ at the cost of increasing the variance of the orthogonal quadrature: at the angle $\phi$ the variance $\Delta J_\phi ^2$ becomes minimal, whereas the orthogonal variance $\Delta J_{\phi+\pi/2} ^2$ becomes maximal \cite{kitagawa:93}. For example, the squeezed state shown in figure~\ref{fig:blochintro} has reduced number fluctuations, as described by the normalized number squeezing factor $\xi_N=\Delta J_z/(\sqrt{N}/2)$, and enhanced phase fluctuations, as described by the normalized phase squeezing factor $\xi_{\rm phase}=\Delta J_y/(\sqrt{N}/2)$.

\begin{figure}
\center\includegraphics[width=0.75\columnwidth]{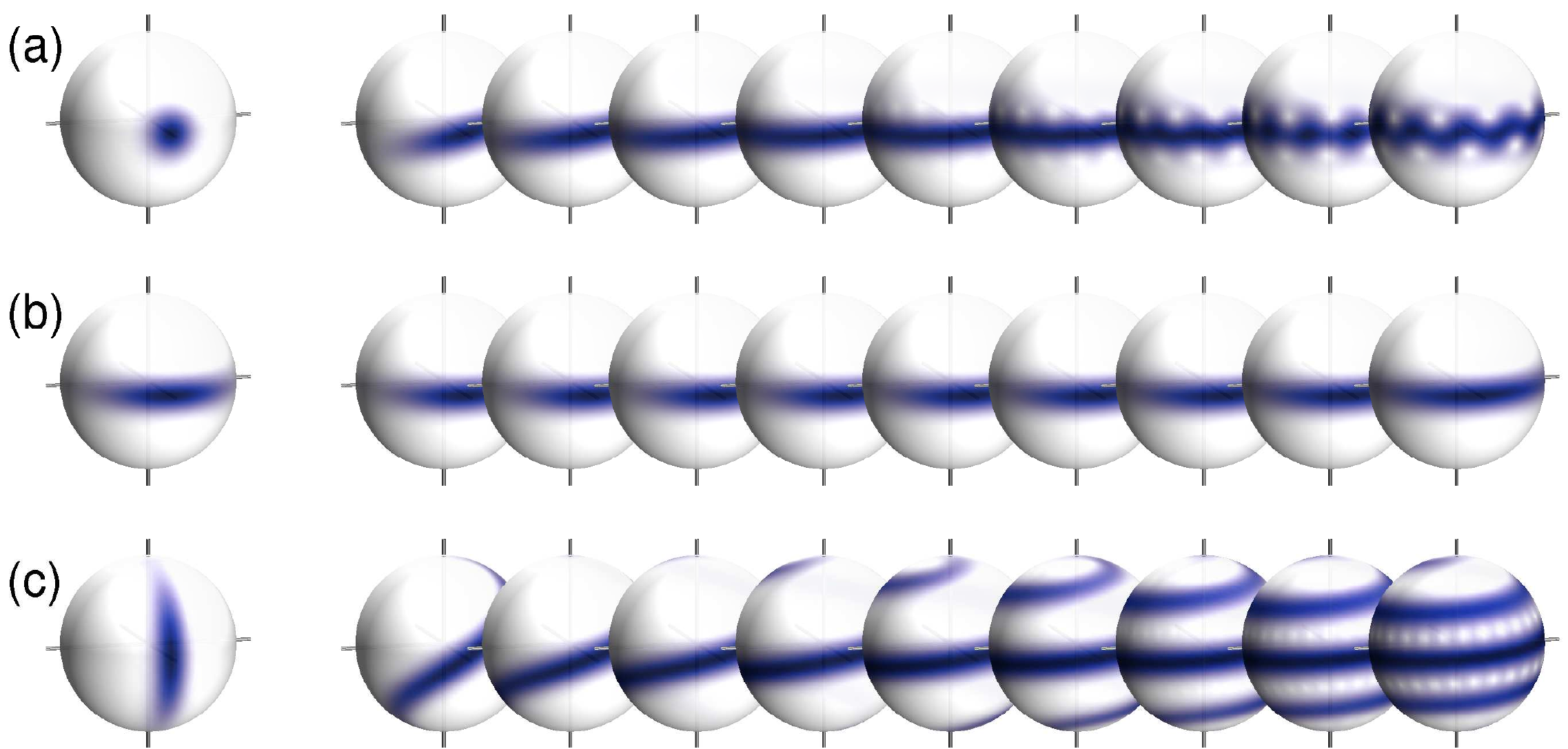}
\caption{Time evolution of states on Bloch sphere. The different panels report results for a (a) binomial, (b) number squeezed [$\xi_N\approx0.2$], and (c) phase squeezed state [$\xi_{\phi}\approx0.2$]. The time interval is $T=12.8$ for the phase squeezed state, and $T=32$ otherwise. Due to the non-linear atom-atom interactions the states become distorted, thus spoiling the interferometer performance.}\label{fig:blochtwist}
\end{figure}

Within the pseudo-spin framework, the two-mode Hamiltonian becomes 
\begin{equation}\label{eq:hamtwomode.pseudospin}
  \hat H=-\Omega\hat J_x+2\kappa \hat J_z^2
\end{equation}
which is completely analogous to the Josephson Hamiltonian of superconductivity. $\Omega$ is associated with the (time-dependent) Josephson energy and the non-linearity $\kappa$ with the charging energy \cite{barone:82}. For a given state on the Bloch sphere, the tunnel coupling rotates the state around the $x$-axis, whereas the non-linear part distorts the state such that the components above and below the equator become twisted to the right- and left-hand side, respectively. The twist rate due to $\hat J_z^2$ increases with distance from the equator (see figure \ref{fig:blochtwist}).

\subsection{Readout noise in the interference pattern} 

We next address the question which states would be the best for the purpose of \textit{reading} out an atom interferometer. For this purpose we neglect the non-linear atom-atom interaction that distorts the wave function during the phase accumulation time and postpone the question of how the specific states could be actually prepared in experiment. Our discussion (which closely follows Ref.~\cite{jaaskelainen:04}, with some extensions) is primarily intended to set the stage for the later discussion of the full atom interferometer sequence in presence of atom-atom interactions. 

We assume that in the phase accumulation stage the wave function has acquired a phase $\theta$. Instead of describing the interaction process dynamically, which would correspond to a Bloch-sphere rotation of the state around the $z$-axis, we directly assign the phase to the single-particle wave function, such that the field operator reads \cite{jaaskelainen:04}
\begin{equation}\label{eq:fieldoptheta}
  \hat\Psi(x)=\hat a_L\,\phi_L(x)+\hat a_R\,e^{-i\theta}\phi_R(x)\,.
\end{equation}
To read out $\theta$, one usually turns off the double well potential and lets the two atom clouds overlap in time of flight (TOF). The accumulated phase is then determined from the interference pattern \cite{schumm:05}.  After release, the wave function evolves under the free Hamiltonian. Here and in the rest of this paper we ignore the influence of atom-atom interactions during TOF which is justified in low dimensional systems \cite{imambekov:09}, but might be problematic under other circumstances \cite{xiong:06,masiello:07,cederbaum:08}. 

As a representative example, we consider for the dispersing wavefunctions $\phi_L(x)$ and $\phi_R(x)$ two Gaussians with variance $\sigma^2$, which are initially separated by the interwell distance $d$. The density operator $\hat n(x)=\hat\Psi^\dagger(x)\hat\Psi(x)$ associated with this TOF measurement can then be computed by using the pseudospin operators of equation~\eqref{eq:pseudospin} as
\begin{equation}\label{eq:densityop}
  \mbox{\hspace*{-2cm}}
  \hat n(x)=\frac N2(n_L+n_R)+2\sqrt{n_Ln_R}\Bigl(
  \cos(kx+\theta)\hat J_x-\sin(kx+\theta)\hat J_y\Bigr)+(n_L-n_R)\hat J_z\,.
\end{equation}
%

In the following we analyze the mean (ensemble-averaged) atomic density
and its ﬂuctuations around the mean value, such that for a single-shot mea-
surement the outcome is with a high probability within the error bonds
deﬁned by the variance if the probability distribution is Gaussian.
 If the initial state preparation was for a symmetric double well potential, then the mean atom number difference vanishes  $\langle \hat J_z\rangle=0$, and for the real-valued tunnel coupling we can set $\langle \hat J_y\rangle=0$. The mean density thus becomes
\begin{equation}\label{eq:denop}
  \mbox{\hspace*{-2cm}}
  n(x)=\left<\hat n(x)\right>=
  \frac N2\Bigl(n_L(x)+n_R(x)\Bigr)+2\sqrt{n_L(x)n_R(x)}\cos(kx+\theta)\left<\hat J_x\right>\,,
\end{equation}
which is shown for representative examples in figure~\ref{fig:den}. Here, the visibility of the interference fringes [equation~\eqref{eq:denop}] is determined by the \textit{coherence factor}
\begin{equation}\label{eq:coherence}
  \alpha=2\left<\hat J_x\right>/N\,.
\end{equation}
It is one for a coherent state with perfect polarization, $\langle \hat J_x \rangle = N/2$, where all atoms reside in the bonding orbital $\phi_L(x)+e^{-i\theta}\phi_R(x)$, while $\langle \hat J_y \rangle = \langle \hat J_z \rangle = 0$. In contrast, a Fock-type state, where half of the atoms reside in the left well and the other half in the right well, has no defined phase relation between the orbitals: it is completely delocalized around the equator of the Bloch sphere, and the coherence $\alpha=0$ vanishes. In general, squeezed states have $0<\alpha<1$.

From equation~\eqref{eq:densityop} we can also obtain the density fluctuations 
\begin{eqnarray}\label{eq:denopfluc}
  &&\mbox{\hspace*{-2.0cm}}
  \Delta n(x)^2=\langle \bigl(\hat n(x)-n(x)\bigr)^2\rangle \\ 
   &&\mbox{\hspace*{-0.7cm}}=
  4n_Ln_R\left(\left<\Delta\hat J_x^2\right>\cos^2(kx+\theta)+
               \left<\hat J_y^2\right>\sin^2(kx+\theta)\right)
  +(n_L-n_R)^2\left<\hat J_z^2\right>\,.\nonumber
\end{eqnarray}
Noise contributions from all pseudospin operators contribute, differently weighted by the time and space dependent probability distributions $n_{L,R}(x,t)$. The $x$-contribution, proportional to $\langle\Delta\hat J_x^2\rangle$, accounts for an uncertainty in the polarization of the state on the equator (see figure~\ref{fig:blochintro}). It is zero only for a coherent state. We will discuss the physical meaning of this quantity later in context of MCTDHB, section~\ref{sec:MCTDHB}. The $y$-contribution accounts for the intrinsic phase width of the quantum state, which provides a fundamental limit for the phase measurement, and the $z$-contribution for the number fluctuations between the two wells.

\subsection{Phase sensitivity}

The ideal state for detecting small variations $\Delta\theta$ is one where $n(x)$ as a function of $\theta$ has a sufficiently large derivative, and the fluctuations $\Delta n(x)$ are sufficiently small. In order to resolve $\Delta\theta$ the inequality 
\begin{equation}\label{eq:sigtonoise}
  \Delta n(x)\le\left|\frac{\partial n(x)}{\partial\theta}\right|\Delta\theta
\end{equation}
has to be fulfilled. The explicit expression in terms of $n_{L,R}$ follows immediately from equations~\eqref{eq:denop} and \eqref{eq:denopfluc}. A particularly simple expression is obtained if we keep in equation~\eqref{eq:sigtonoise} only the dominant contribution from the phase noise, which is proportional to $\langle\hat J_y^2\rangle$ (other contributions depend on $n_{L,R}(x)$ and are expected to be less important), which leads to the  estimate for the phase sensitivity in terms of shot noise by Kitagawa and Ueda \cite{kitagawa:93} (see also table \ref{tab:squeezing})
\begin{equation}\label{eq:phasesqueezing}
  \xi_R=\frac{\sqrt{N\langle\Delta\hat J_y^2\rangle}}{\left|\langle\hat J_x\rangle\right|}\,.
\end{equation}
This expression will serve us as a guiding principle for optimizing atom interferometry. We will consider in the following the phase squeezing at the optimal working point. The normalization by $\langle \hat J_x \rangle$ takes into account that improving interferometric sensitivity requires not only to reduce noise but also to maintain a high interferometer contrast $\alpha$ (which determines the difference to $\xi_{\rm phase}$ via $\xi_R=\xi_{\rm phase}/\alpha$) . Equation~\eqref{eq:phasesqueezing} provides a simple way to estimate the sensitivity obtainable by a given initial state. For a coherent state with a binomial atom number distribution, the mean value and the variance are proportional to the total atom number $N$. Thus, $\xi_R=1$ and the phase sensitivity $\Delta\theta=1/\sqrt N$ is {shot-noise} limited. We refer to this limit as the \textit{standard quantum limit}.

\begin{table}

\begin{center}
\begin{tabular}{l|l}

Squeezing factor & Definition \\
\hline
\hline
Useful squeezing & $\xi_R=\Delta J_y/(\alpha\sqrt{N}/2)$ \\
Phase squeezing & $\xi_{\rm phase}=\Delta J_y/(\sqrt{N}/2)=\xi_R\alpha$ \\
Number squeezing & $\xi_N=\Delta J_z/(\sqrt{N}/2)$ \\
\end{tabular}
\end{center}
\caption{Definition of squeezing factors used in this work. $\xi_R$ is the useful squeezing that determines the sensitivity of an interferometer. It differs from the phase squeezing through the coherence factor $\alpha=2\langle \hat J_x \rangle/N$. $\xi_N$ is the number squeezing that depends on the number fluctuations between the two wells.
\label{tab:squeezing}}
\end{table}

For the purpose of reading the interference pattern, phase squeezed states are ideal because they allow to reduce the sensitivity below shot noise. $\xi_R$ provides a measure of \emph{useful squeezing} for metrology \cite{wineland:94}: a state with $\xi_R < 1$ allows to overcome the standard quantum limit by a factor $\xi_R$. The lower bound of the sensitivity is provided by the Heisenberg limit. From the commutation relation $[\hat J_y,\hat J_z]=i\hat J_x$ one obtains the uncertainty relation $\Delta J_y\Delta J_z\ge N/2$. Thus, for a state with a maximal number uncertainty, $\Delta J_z=N/2$, the standard deviation $\Delta J_y$ is on the order of unity, and the Heisenberg limit becomes $\xi_R=\sqrt{2/N}$.

\begin{figure}
\includegraphics[width=\columnwidth]{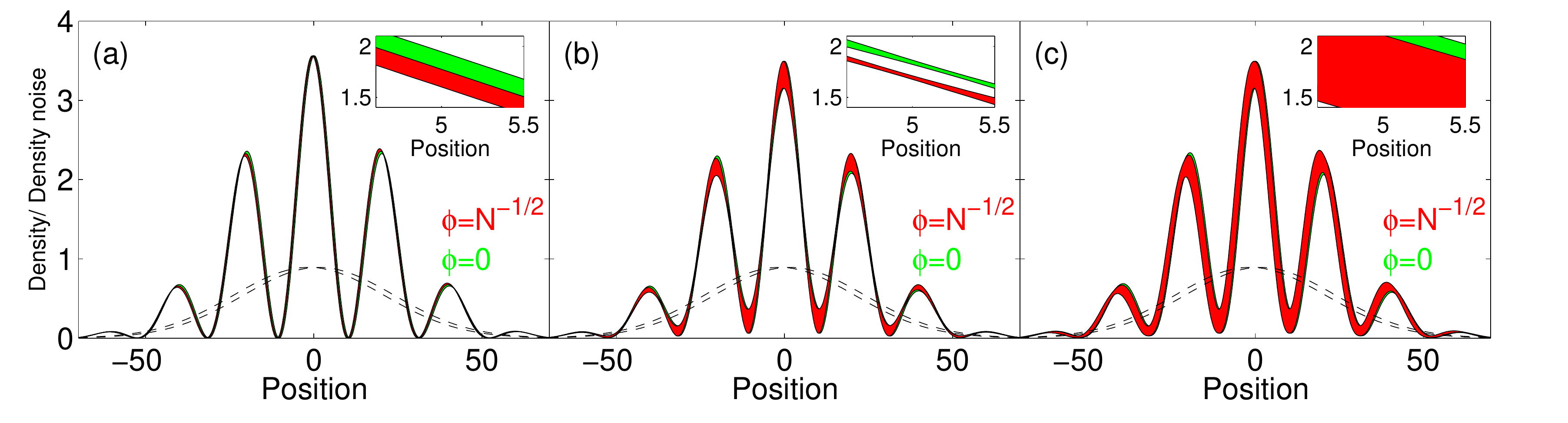}
\caption{Density and density noise for (a) binomial, (b) phase squeezed, and (c) number squeezed state, and for a phase $\theta=0$ (green areas) and $\theta=1/\sqrt{N}$ (red areas), corresponding to shot noise. We use $N=100$, width $\sigma=0.05$, a read-out time $T_{\rm tof}=10$ and an interwell separation of $d=5$. The degree of squeezing is characterized by a squeezing factor of $\xi_N=0.18$ ($\xi_{\phi}=0.18$). The dashed lines show the density profile of the condensates without interference. The insets magnify the region where the noise is least. The phase sensitivity is best for the phase squeezed state, although in certain regions noise is enhanced due to $\Delta J_x$. \label{fig:den}}
\end{figure}

In figure~\ref{fig:den} we show density and density fluctuations for a (a) binomial, (b) phase-squeezed, and (c) number squeezed state. One observes that the fluctuations are smallest for the phase-squeezed state, and become larger for the coherent and number squeezed state.  To appreciate the sensitivity of the interferometer, we also plot the density profile for a state that has acquired a phase $\theta$ of the order of shot noise. As apparent from the insets, which magnify the regions of smallest noise, the smallest $\theta$ variations can be resolved with the phase-squeezed state. 

In a typical double well interferometer experiment, where the interference is read out in TOF, the exact number of atoms $N$ is not known, and one has to obtain $N$ and the accumulated phase $\theta$ from a suitable fitting procedure. To benefit from the regions of reduced density noise, the density distribution $n(x)$ has to be weighted appropriately by $\Delta n(x)$. 

\begin{figure}[b]
\includegraphics[width=\columnwidth]{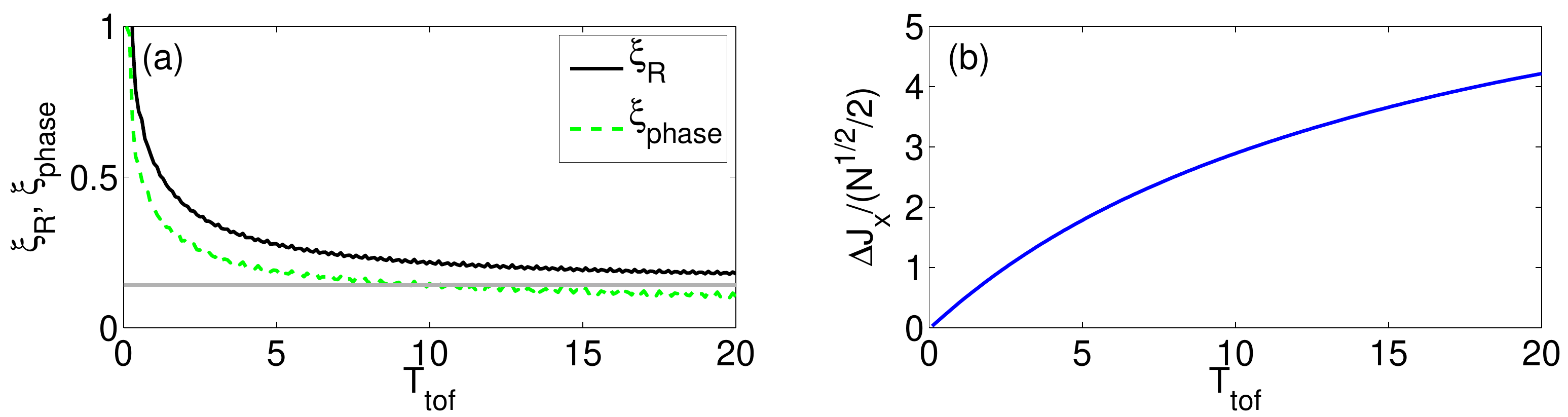}
\caption{(a) Best achievable $\xi_R$ versus read-out time $T_{\rm tof}$ for atom number $N=100$, $\sigma=0.05$, and $d=5$ (as in Fig.~\ref{fig:den}). For given $T_{\rm  tof}$, the optimal phase squeezing $\xi_{\rm phase}$ (dashed line) is found such that $\xi_R$ (solid black line) is minimized. $\xi_R$ approaches the Heisenberg limit (gray line) for large $T_{\rm tof}$. The difference between $\xi_R$ and $\xi_{\rm phase}$ is due to $\alpha < 1$ (b) Polarization noise, $\Delta J_x$, of the corresponding phase squeezed states of (a). 
\label{fig:den2}}
\end{figure}

The dependence on the duration of the read-out stage is shown in figure~\ref{fig:den2}. For short $T_{\rm tof }$ the overlap between the condensates is small, and the contribution from $\Delta J_z$ (the third term in equation~\eqref{eq:denopfluc}) important.  Since small phase fluctuations are accompanied by large number fluctuations, there exists an optimal degree of phase squeezing for the interferometer input state. For large enough $T_{\rm tof}$ this effect is less important and $\xi_R$ approaches the Heisenberg limit. We optimize the degree of phase squeezing of the initial state (i.e., at $T_{\rm tof}=0$) for given $T_{\rm tof}$, such that $\xi_R$ is minimized. This is shown in (a) versus $T_{\rm tof}$ (solid line), together with the corresponding optimal phase squeezing $\xi_{\rm phase}$ (dashed line).   For small $\xi_{\rm phase}$, polarization noise $\Delta J_x$ [shown in panel (b)] is introduced. This noise reduces the spatial region where the sensitivity is high [see also figure~\ref{fig:den}(b)]. In conclusion, the expansion in TOF has to be sufficiently long, such that the atom clouds can expand sufficiently far, and the phase sensitivity is no longer limited by number fluctuations between the clouds.

\subsection{Interferometry in presence of atom-atom interactions during the phase accumulation stage \label{subsec:interactions}}

In the following we discuss how the atom-atom interactions affect the phase distribution of the split condensate during the phase accumulation stage and spoil atom interferometry, and what could be done to minimize those effects.

For the purpose of atom interferometry, it is convenient to introduce the \textit{phase eigenstates} \cite{menotti:01} 
\begin{equation}
  |\phi\rangle=\frac{1}{\sqrt{2\pi}}\sum_m e^{i\phi m}|m\rangle\,,
\end{equation}
where $|m\rangle$ is a state with atom number imbalance $m$ between the left and right well. The relation between $|m\rangle$ and $|\phi\rangle$ corresponds to a Fourier transformation. We can now project any state on the phase eigenstates and obtain the phase representation of the state. From this we obtain the
 phase width $\Delta \phi$ (see also \cite{menotti:01}). 
The noise $\Delta J_y$, which enters the phase sensitivity of the TOF-interferometer according to equation~\eqref{eq:phasesqueezing}, has a similar time evolution as the phase width . However, while the phase width is bound to values below $2\pi$, the upper bound of $\Delta J_y$ is given by the total atom number $N$, and therefore $\Delta J_y=\sqrt{N}\Delta \phi$. 

Phase diffusion due to atom-atom interactions is best illustrated in the Bloch sphere representation (see figure~\ref{fig:blochtwist}): The non-linear coupling $\kappa\hat J_z^2$ from equation~\eqref{eq:hamtwomode.pseudospin} twists the state, and the larger the number fluctuations $\Delta J_z$ the faster the state winds around the Bloch sphere. In the notion of phase eigenstates, a state with a well defined phase has a broad atom-number distribution. As each atom-number eigenstate $|m\rangle$ evolves with a different frequency, the time evolution of a superposition state will suffer phase diffusion. The phase width broadens with rate \cite{javanainen:97}
\begin{equation}\label{eq:diffrate}
  R= 8U_0\Delta n=8U_0\xi_N(\sqrt N/2)\,.
\end{equation}
As a result, states with small number fluctuations ($\xi_N <1$) are more stable during the phase accumulation time.  

One immediately sees a conflict of requirements.  For best readout of the interference pattern we want phase squeezed states, but those are very fragile and result in a fast phase diffusion and a short measurement time.  On the other hand, number-squeezed states allow for longer measurement times but have a rather poor readout performance.  In the remaining part of the paper we will discuss the optimal strategy for interference experiments, and how one can implement them in realistic settings.
 
\begin{figure}
\center\includegraphics[width=0.7\columnwidth]{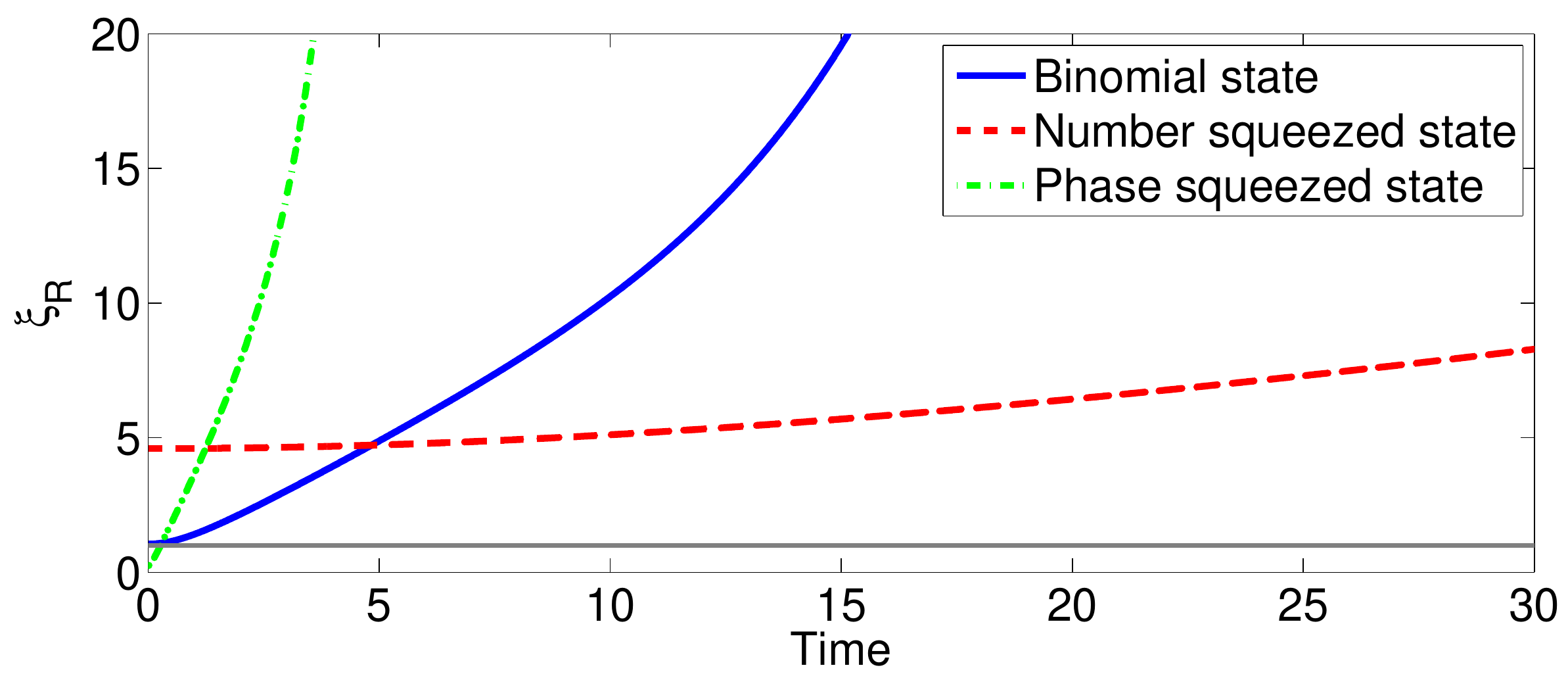} 
\caption{Example for the phase sensitivity of a binomial (solid line), a number squeezed (dashed line), and a phase squeezed (dashed-dotted line) state with squeezing factors $\xi_N\approx 0.22$ and $\xi_{\phi}\approx 0.22$, respectively. Parameters are $N=100$ and $U_0 N=1$. The binomial and phase squeezed states have a better initial phase sensitivity, whereas the number squeezed is much more stable against phase diffusion, and has a much better sensitivity at later times. The gray line corresponds to shot noise ($\xi_R=1$).
\label{fig:demo}}
\end{figure}

To demonstrate the above reasoning, we show in figure~\ref{fig:demo} the achievable phase sensitivity as a function of phase accumulation time $T_{\rm phase}$ in presence of atom-atom interactions for a binomial (solid), a number squeezed (dashed line), and a phase squeezed (dashed-dotted line) state with squeezing factors $\xi_N\approx 0.22$ and $\xi_{\phi}\approx 0.22$, respectively. The phase squeezed state has initially sub-shot noise phase sensitivity. For longer hold times, the number squeezed state outperforms the phase squeezed and binomial ones due to its smaller phase diffusion rate. 

\section{Optimizing atom interferometry\label{sec:opt}}

When designing a trapped atom interferometer for measurements, one has to consider the conflicting requirements from phase diffusion and readout, the one asking for number squeezing, the other for phase squeezing.  In the following we will outline a few strategies of how to optimize interferometer performance for realistic double well settings.

\subsection{A very simple estimate}\label{sec:opt.simple}

We first analyze the impacts of trap geometry and initial state preparation. We start by employing a simple model to illustrate the effects of the non-linearity in the evolution of the split trapped BEC. Let us first look at the interaction energy of a trapped cloud, and how it changes with the number of trapped particles.  Without loosing generality, we discuss in the context of a harmonic trap characterized by the mean confinement $\omega_0$ and the length scale $a_0$, which are defined through
\begin{equation}
 \omega_0 = \sqrt[3]{\omega_x \omega_y \omega_z}\,, \qquad  
 a_0=\sqrt{\frac{\hbar}{m \omega_0}} \, .
\end{equation}
With $N$ atoms in the trap the chemical potential $\mu$ in Thomas-Fermi approximation \cite{dalfovo:99} is given by
\begin{equation}
  \mu = \frac{\hbar \omega_0}{2} \left( \frac{15 \, N \, a_{s}}{a_0} \right)^{\frac{2}{5}}\,,
\end{equation}
where $a_s$ is the $s$-wave scattering length.
Adding a single atom to the trap, changes $\mu$ by 
\begin{equation}	
  \frac{\partial \mu}{\partial N} = \frac{2}{5} \frac{\mu}{N}\,.
\end{equation}
This quantity corresponds to the effective 1D interaction parameter $U_0$ discussed earlier. We can now estimate the scaling of the phase diffusion rate $R$ caused by a number distribution with fluctuations $\Delta N = \xi_N \sqrt{N}/2$ after splitting ($\xi_N=1$ corresponds to a binomial number distribution, whereas  $\xi_N < 1$ to number squeezing),
\begin{equation}
  R \propto \xi_N \, N^{-\frac{1}{10}} \, 
  \omega_0 ^{\frac{6}{5}} \, a_{s} ^{\frac{2}{5}} \, m^{\frac{1}{5}} \,.
\end{equation} 
For $^{87}$Rb atoms ($a_{s}=5.2$ nm, $m=87$) one obtains $R \approx 0.022 \, \xi_N \, N^{-\frac{1}{10}} \,  \omega_0 ^{\frac{6}{5}}$ s$^{-1}$, or in scaled units $R \approx 0.29 \, \xi_N \, N^{-\frac{1}{10}} \,  \omega_0 ^{\frac{6}{5}}$.
If we now set the phase diffusion rate $R$ equal to the phase accumulation rate $\frac{1}{\hbar}\Delta E$ (the signal we want to measure), we find the limit $\frac{1}{\hbar} \Delta E^{\rm min} = R$ for the sensitivity of a single-shot interferometer measurement, even for perfect readout. It is interesting to note that this sensitivity limit is only very weakly dependent on the atom number $N$. 

If the interferometer measurement is not limited by readout, we can identify the following strategies for improving the interferometer performance (or, equivalently, reducing the effect of phase diffusion).
\begin{description}
	\item[Minimize the scattering length.] The best is to set $a_{s}=0$, which can in principle be achieved by employing Feshbach resonances \cite{chin:10}.  Drastic reduction of phase diffusion when bringing the scattering length close to zero was recently demonstrated in two experiments \cite{gustavsson:08,Fattori:08}. A disadvantage thereby is that using Feshbach resonances requires specific atoms and specific atomic states.  These states need to be tunable, and are therefore \textit{not} the '\textit{clock}' states usually used in precision experiments which are immune to external disturbances like magnetic fields. 
	
	\item[Choose a trap with weak confinement.]  This route seems problematic, since the timescale in splitting and manipulating the trapped atoms scales with the trap confinement.  Optimal control techniques, like discussed in Refs.~\cite{hohenester.pra:07,grond.pra:09,grond.pra:09b}, will be needed to allow splitting much faster than the phase diffusion time scale. It is interesting to note that one needs a strong confinement only in the splitting direction. In the other two space directions the confinement can be considerably weaker. This suggests to work with strongly anisotropic traps. \\
  For an \textit{elongated cigar-shaped} trap (1D geometry) with confinement ratio $C_{1D}=\frac{\omega_z}{\omega_\perp}$ (strong confinement $\omega_\perp$ in radial directions, weak confinement $\omega_z$ in  axial direction) one finds  
\begin{equation}
  \omega_0 ^{(1D)}=\omega_\perp \sqrt[3]{C_{1D}} \, .
\end{equation}
For a \textit{flat pan cake-shaped} trap (2D geometry) with confinement ratio $C_{2D}=\frac{\omega_{\rm plane}}{\omega_\perp}$ (strong transverse confinement $\omega_\perp$ and weak in-plane confinement $\omega_{\rm plane}$) one finds
\begin{equation}
  \omega_0 ^{(2D)}=\omega_\perp \sqrt[3]{C_{2D} ^2} \,.
\end{equation}
With a confinement ratio $C \sim 1/1000$, which is easily obtainable in experiments the phase diffusion is reduced by a factor 10 in a 1D geometry and a factor 100 in a 2D geometry.

\item[Increase number squeezing in the splitting process.] This directly reduces the phase diffusion rate and hence leads to a better limit for the minimal detectable signal. Number squeezing can be achieved during the splitting process.  It is mediated by the atom interactions and one has to achieve a careful balance between the interactions necessary to obtain sizable number squeezing and the decremental effect of the interactions during the phase accumulation time. This will be one of the central parts in our optimization discussed below.
	
\end{description}

In an ideal interferometer one would like to use clock states, create strong squeezing during the splitting process, exploiting the non-linearity in the time evolution, and then turn off the interactions (by setting the scattering length to $a_s = 0$) after splitting.  All together might, however, be difficult or even impossible to achieve. In the remainder of the manuscript we will discuss the different contributions to the precision of an atom interferometer, and investigate how the performance can be optimized.

\subsection{Optimization of the many-boson states\label{subsec:results}}

\begin{figure}
\center\includegraphics[width=0.7\columnwidth]{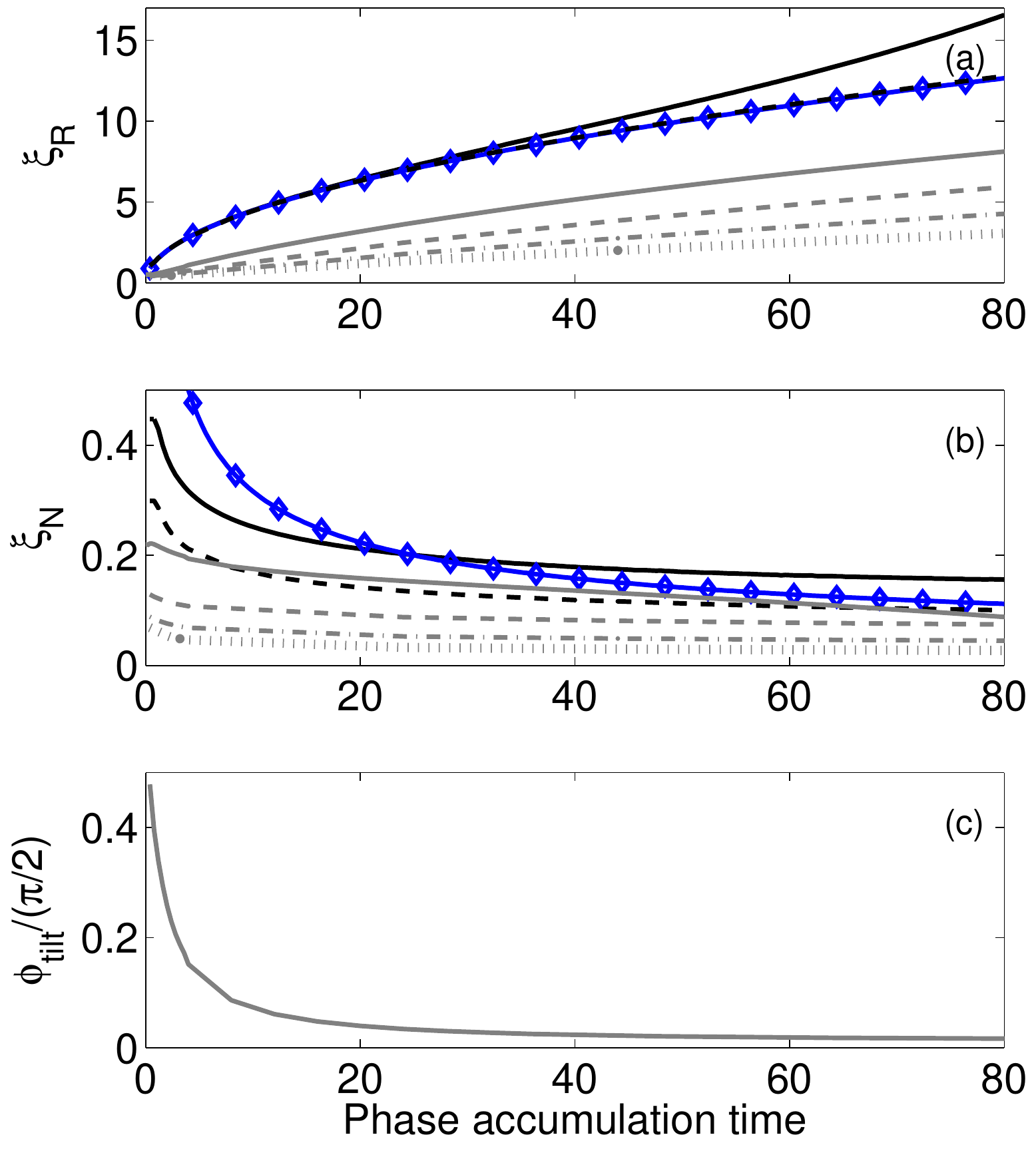}
\caption{(a) Optimal phase sensitivity versus phase accumulation time for $N=100$ (solid lines), $N=500$ (dashed lines), $N=2000$ (dashed-dotted lines), and $N=8000$ (dots). Interaction strength is such that $U_0 N=1$. The black lines show results for number-squeezed states, and the gray lines result for slightly tilted number-squeezed states. (b) The corresponding optimal number squeezing. (c) The total tilt angle $\phi_{\rm tilt}=\Omega_{\rm tilt}T_{\rm pulse}$ is determined by the pulse duration (here $T_{\rm pulse}=2$) and strength $\Omega_{\rm tilt}$ (very similar for all $N$). The diamond symbols in (a) and (b) show estimates from a simple model. 
\label{fig:figxi}}
\end{figure}

\subsubsection{Optimized Number Squeezing \label{subsubsec:optsplit}}

First we shortly discuss how number-squeezed states can be prepared during the splitting stage. Number squeezed states are created in a double well potential when the interaction energy starts to dominate over the tunnel coupling, the latter being controlled by the barrier height and the double well separation.  A natural way to achieve high number squeezing is dynamic splitting of a BEC \cite{menotti:01,streltsov:07}, such that the wavefunction can adiabatically follow the ground state \cite{javanainen:99}. However, this may take very long, possibly longer than the phase diffusion time of the split condensate.  In our earlier work \cite{grond.pra:09,grond.pra:09b,hohenester.fdp:09} we employed optimal control theory (OCT) to find splitting protocols which allow for high number squeezing on a fast timescale, at least one order of magnitude shorter than for the quasi-adiabatic splitting.   These protocols can be viewed as the continuous transformation of a (close to) harmonic potential into a double well. Thereby, a barrier is ramped up at the center, and simultaneously the two emerging wells are separated. In many cases this splitting process can be parameterized by a single parameter, whose time variation is obtained within optimal control theory such that sizeable squeezing is created, condensate oscillations are prevented after splitting, and phase coherence is better preserved at the end of the splitting process \cite{grond.pra:09,grond.pra:09b,hohenester.fdp:09}.

Unless stated otherwise, in the following discussion of the dynamics during the phase accumulation stage we use number-squeezed states as initial states which are obtained as the ground states of equation~\eqref{eq:hamtwomode.pseudospin} for finite values of tunneling $\Omega$. They are very similar to those obtained by OCT the splitting. Similar initial states obtained by exponential splitting have a smaller degree of coherence. During the phase accumulation stage, we set $\Omega=0$.

How much number squeezing is ideal for a pre-determined phase accumulation time of an interferometer? In figure~\ref{fig:figxi} we show results where we optimize the degree of number squeezing for the interferometer input states at $T_{\rm phase}=0$, in order to achieve the best phase sensitivity at a given time $T_{\rm phase}$. The top panel reports the best achievable phase sensitivity for an initially number squeezed state (black lines), and the middle panel reports the corresponding number squeezing. For short phase accumulation times, less initial number squeezing is better. With increasing accumulation time more number squeezing becomes favorable. This is due to the competition between phase fluctuations $\Delta J_y$, which increase with number squeezing, and the decrease of phase diffusion for states with high number squeezing. 

The results can be well explained by a simple model. Neglecting the effects of reduced phase coherence, we have initially $\xi_R=\xi_{\rm phase}^0$, the initial phase squeezing. Phase diffusion with rate $R$ then results in $\xi_R=\sqrt{(\xi_{\rm phase}^0)^2+R^2 T_{\rm phase}^2}$. We next use that $\xi_N \xi_{\rm phase}^0\approx 1$, which is in the spirit of the Heisenberg uncertainty principle and agrees well with our OCT results. Putting in all the constants, we have
\be
\xi_R= \sqrt{\frac{1}{\xi_N^2 N}+16 N\xi_N^2U_0^2 T_{\rm phase}^2}. 
\ee
The minimum with respect to $\xi_N$ is found as $\xi_N^{\rm min}=1/\bigl(2\sqrt{U_0 N T_{\rm phase}}\bigr)$, which yields a best phase sensitivity $\xi_R^{\rm min}=2\sqrt{2U_0 N T_{\rm phase}}$. For a given final $\xi_R$ we see that $T_{\rm phase}$ is indirectly proportional to the interaction parameter $U_0$, which allows rescaling of $T_{\rm phase}$ in case of a different $U_0$. 

Predictions of the simple model are shown in figure~\ref{fig:figxi} by the diamond symbols. The agreement with the exact results is very good in (a), and gives the right scaling in (b). Indeed, $\xi_R^{\rm min}$ in figure~\ref{fig:figxi} is independent of $N$ for long times, as long as $U_0 N$ is constant. This is not true for $\xi_N^{\rm min}$. The reason is that neglecting phase coherence makes the minima with respect to $\xi_N$ much more shallow. For longer times, however, phase coherence becomes more important and the present approximations are no longer valid for small N.

\subsubsection{Optimized specialized initial states \label{subsubsec:spec}}

A different strategy for improving the interferometry performance is to prepare the system at the beginning of the phase accumulation stage (i.e., at $T_{\rm phase}=0$) in a special state, which evolves under the influence of the non-linear interaction after some pre-determined time into a state with high intrinsic sensitivity. The ideal initial state would be the time reversal of a phase squeezed state. We denote this strategy as \textit{refocusing}. When the condensate is released at the optimal time, and expands in absence of interactions to form the interference pattern, interferometry can be performed with a sensitivity determined by the properties of the refocused state. This can be achieved because the phase accumulation (rotation around $z$-axis on the Bloch sphere) and the non-linear coupling do not interfere. 

Such a refocusing strategy is related to spin echo techniques, which where investigated by turning the scattering length $a_s$ from repulsive to attractive \cite{widera:08}. However, the latter has given a rather poor improvement, because it does not lead to a perfect time reversal of the many-body dynamics \cite{sakmann2:09}. Artificial preparation of the desired time reversed states, seems to be very difficult. We did not succeed in this task using optimal control techniques.

\begin{figure}
\begin{tabular}{ll}
\includegraphics[height=0.37\columnwidth]{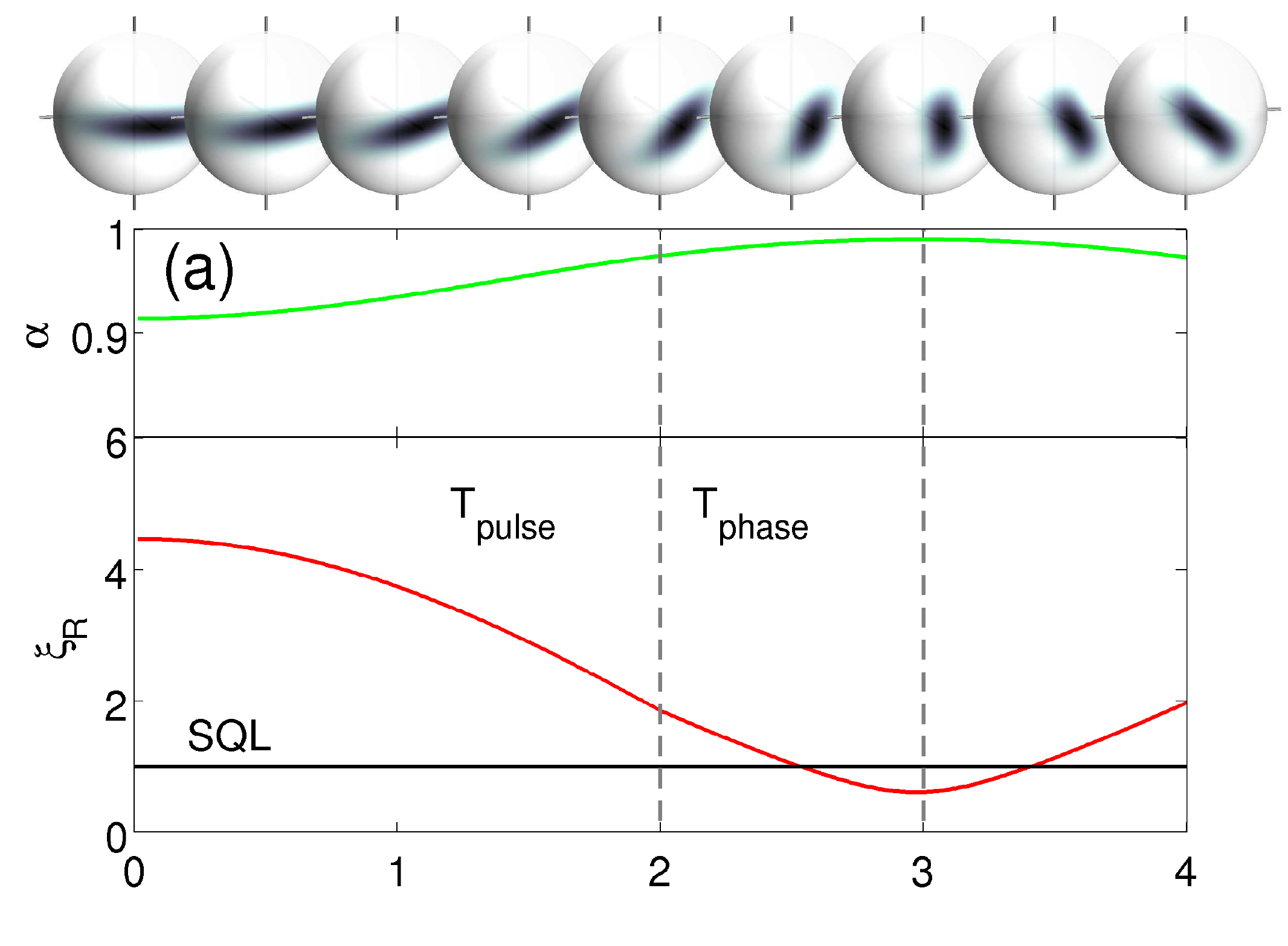}&\includegraphics[height=0.37\columnwidth]{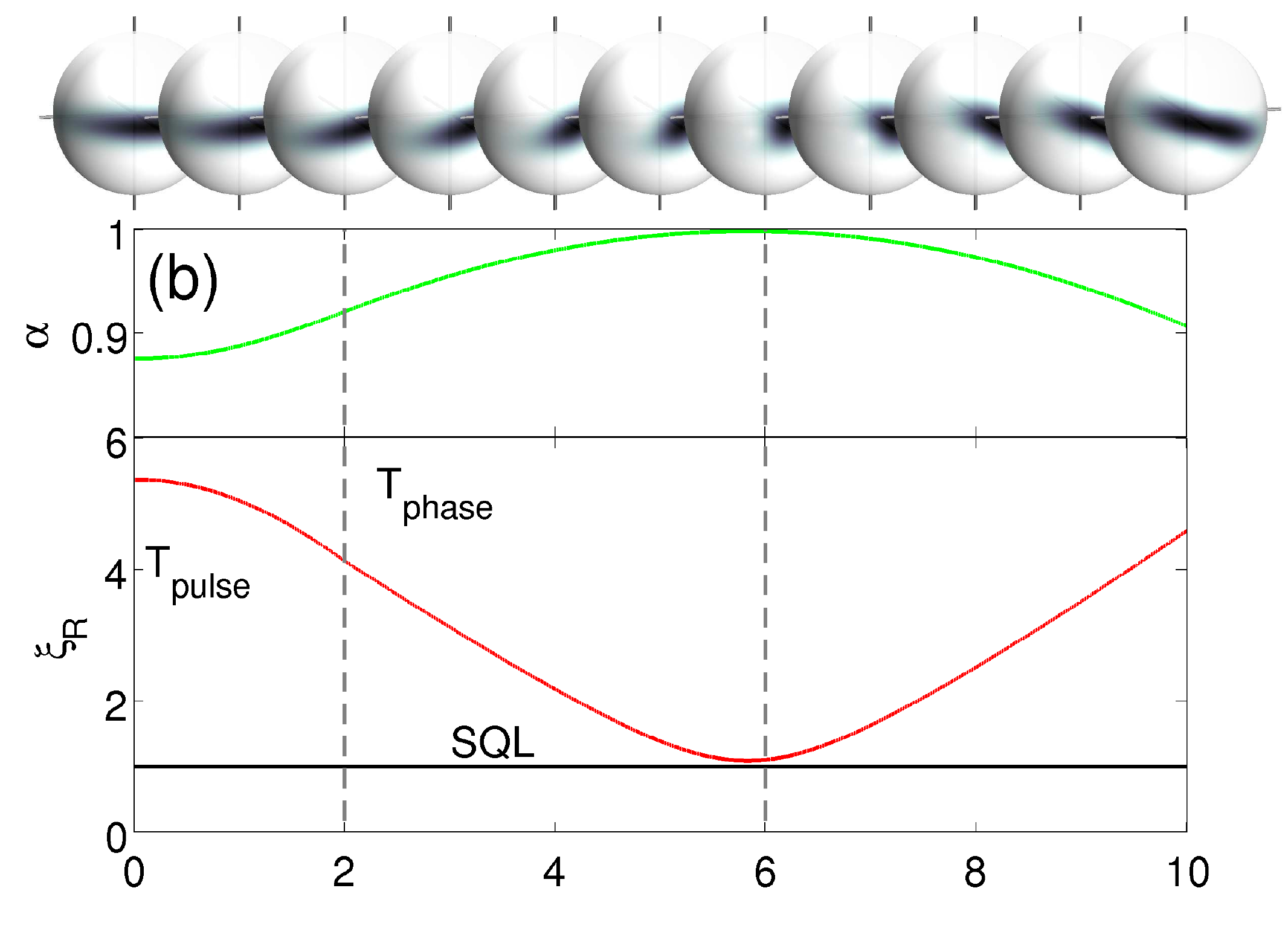}
\end{tabular}
\caption{ Refocusing control sequence for $N=100$, $U_0 N=1$, $T_{\rm pulse}=2$, and (a) $T_{\rm phase}=1$ as well as (b) $T_{\rm phase}=4$. The lower panel shows $\xi_R$, while the upper panel the coherence factor $\alpha$. In the first stage, the tunnel pulse  is applied and \emph{tilting} is achieved. In the second stage, the state refocuses to a state with a good phase sensitivity, in (a) below shot-noise. The Bloch spheres visualize the time evolution. 
\label{fig:tiltdetails}}
\end{figure}

One state that leads to very good refocusing can be prepared by \textit{tilting} the initial number squeezed state on the Bloch sphere slightly \textit{against} the direction of the twist originating from $\hat J_z^2$. The tilt can be achieved by applying a short tunnel pulse within a time interval $T_{\rm pulse}$ that rotates the number squeezed state.  In real space, this operation corresponds to lowering the barrier for a suitable amount of time, which cannot be done arbitrarily fast because of condensate oscillations. Appropriate controls of the barrier will be discussed in detail in context of a realistic modeling in section~\ref{sec:MCTDHB}. Within the generic model we consider for simplicity square-$\Omega$ pulses, which is the best possible pulse in presence of interactions \cite{pezze2:06}. Examples are shown in figure~\ref{fig:tiltdetails} for different $T_{\rm phase}$. During the \emph{tilting} pulse sequence and the phase accumulation stage, 'rephasing' happens and the phase fluctuations decrease. Simultaneously, the phase coherence is restored to a value close to one.  This significantly improves the phase sensitivity, for short times one can even reach below shot noise. However, this cannot be done perfectly, the degree of phase squeezing $\xi_{\rm phase}$ achieved after refocusing is always less than the degree of number squeezing $\xi_N$ of the original state. 

We next optimize systematically both parameters of initial number squeezing and tilt angle. The lowest panel of figure~\ref{fig:figxi} shows the optimal tilt angles. From the upper panels (bright lines) we find a clear improvement of phase sensitivity for a given phase accumulation time. The dependence on $N$ is very distinct now for small atom numbers, and saturates for large $N$. We find that for small $N$ the improvement is roughly a factor of three in time, and for large $N$ approximately an order of magnitude.

\subsection{Optimize trapping potential and atom number. esults of generic two-mode model\label{subsec:opt_trap}}

We next proceed to a more detailed analysis of the ideal trap parameters. Considering our previous discussion of section~\ref{sec:opt.simple}, we expect an improvement of the interferometer performance when increasing the anisotropy of the trap. In the following calculations we choose a fixed $N=100$ (the results are not expected to depend decisively on $N$). 

\begin{figure}
  \begin{tabular}{l l}
 \includegraphics[width=0.44\columnwidth]{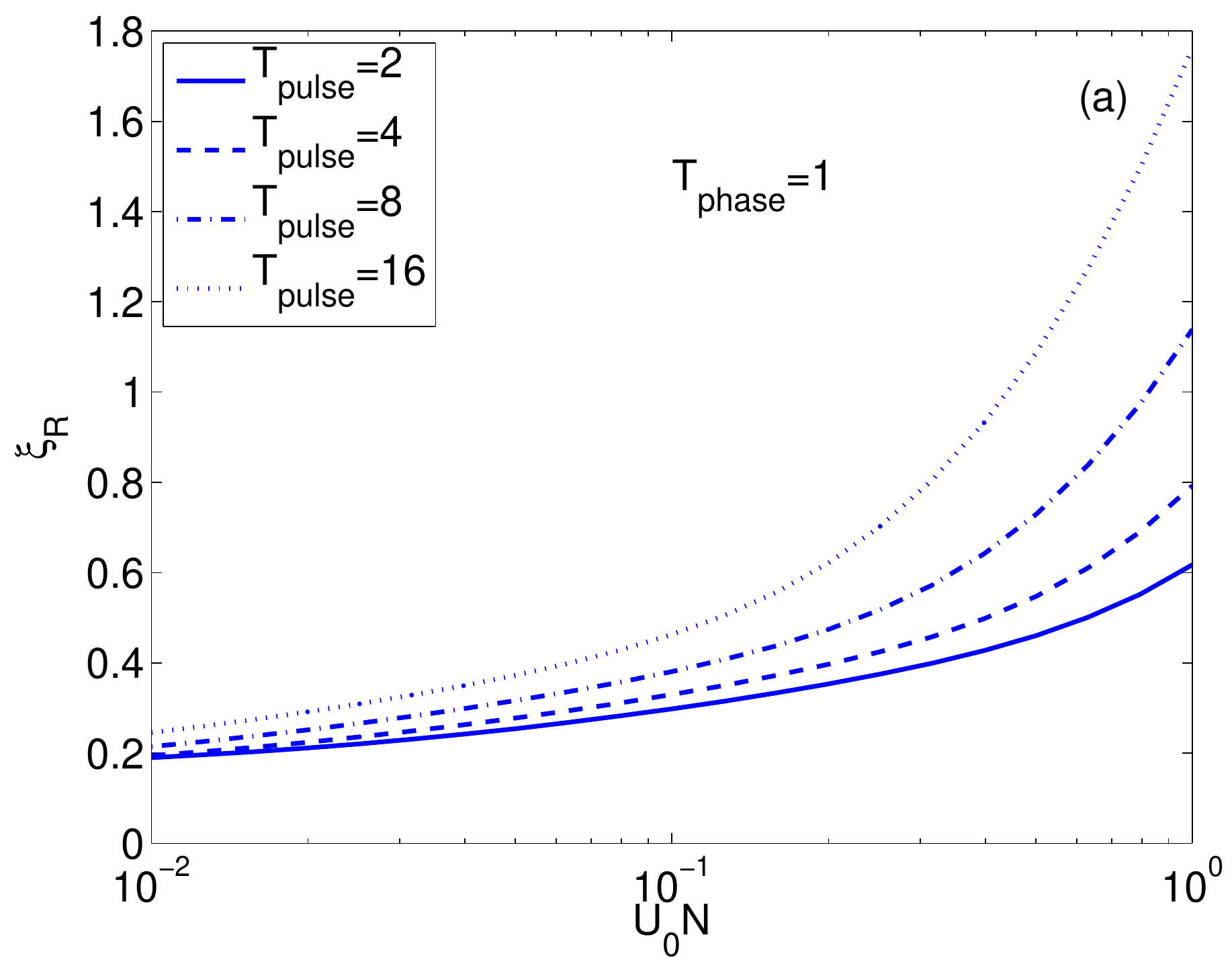}&\includegraphics[width=0.44\columnwidth]{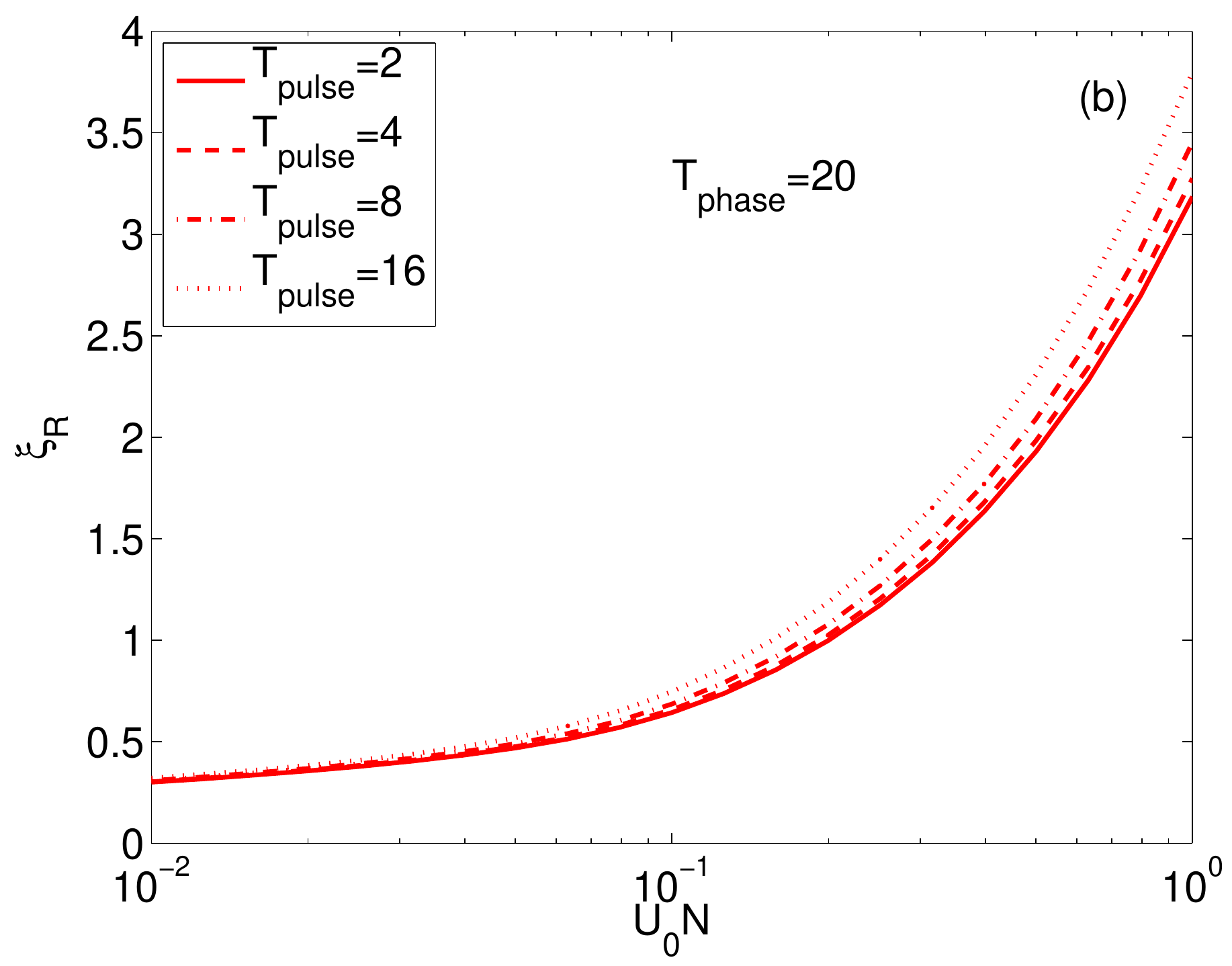}
  \end{tabular}
\caption{ Best achievable phase sensitivity versus interaction strength for various pulse durations for (a) $T_{\rm phase}=1$ and (b) $T_{\rm phase}=20$. Atom number is $N=100$. We optimize for the initial number squeezing and the tunnel pulse. 
 \label{fig:Tpulse}}
\end{figure}

We first investigate the role of the interaction $U_0 N$, and the pulse duration $T_{\rm pulse}$ for the refocusing strategy. In figure~\ref{fig:Tpulse} we plot the best phase sensitivity using refocusing versus interaction strength for (a) $T_{\rm phase}=1$ and (b) $T_{\rm phase}=20$. Sub-shot noise phase sensitivity is clearly achievable for short $T_{\rm pulse}$, or for $U_0 N \ll 1$. Short pulses are favorable and give better phase sensitivity. This is in particular important for short phase accumulation times $T_{\rm phase}$ (see figure~\ref{fig:figxi}). Optimizing the pulse form and duration within a realistic modeling of \emph{tilting} on the Bloch sphere will be discussed in detail in section~\ref{sec:MCTDHB}. 

Quite generally, we can expect that for a reduced interaction parameter it is more difficult to obtain high number squeezing in the splitting process. To estimate the time scale for achieving a certain degree of number squeezing, we consider splitting protocols derived in a previous work within the framework of optimal control theory \cite{grond.pra:09b}, and discussed already in section~\ref{subsubsec:optsplit}. They directly provide us optimized number squeezing for a given splitting time $T_{\rm split}$. For a given phase accumulation time $T_{\rm phase}$, we optimize then the tunnel pulse which tilts the number squeezed state, similar to the analysis of section~\ref{subsubsec:spec}.
\begin{figure}
  \begin{tabular}{l l}
   \includegraphics[width=0.49\columnwidth]{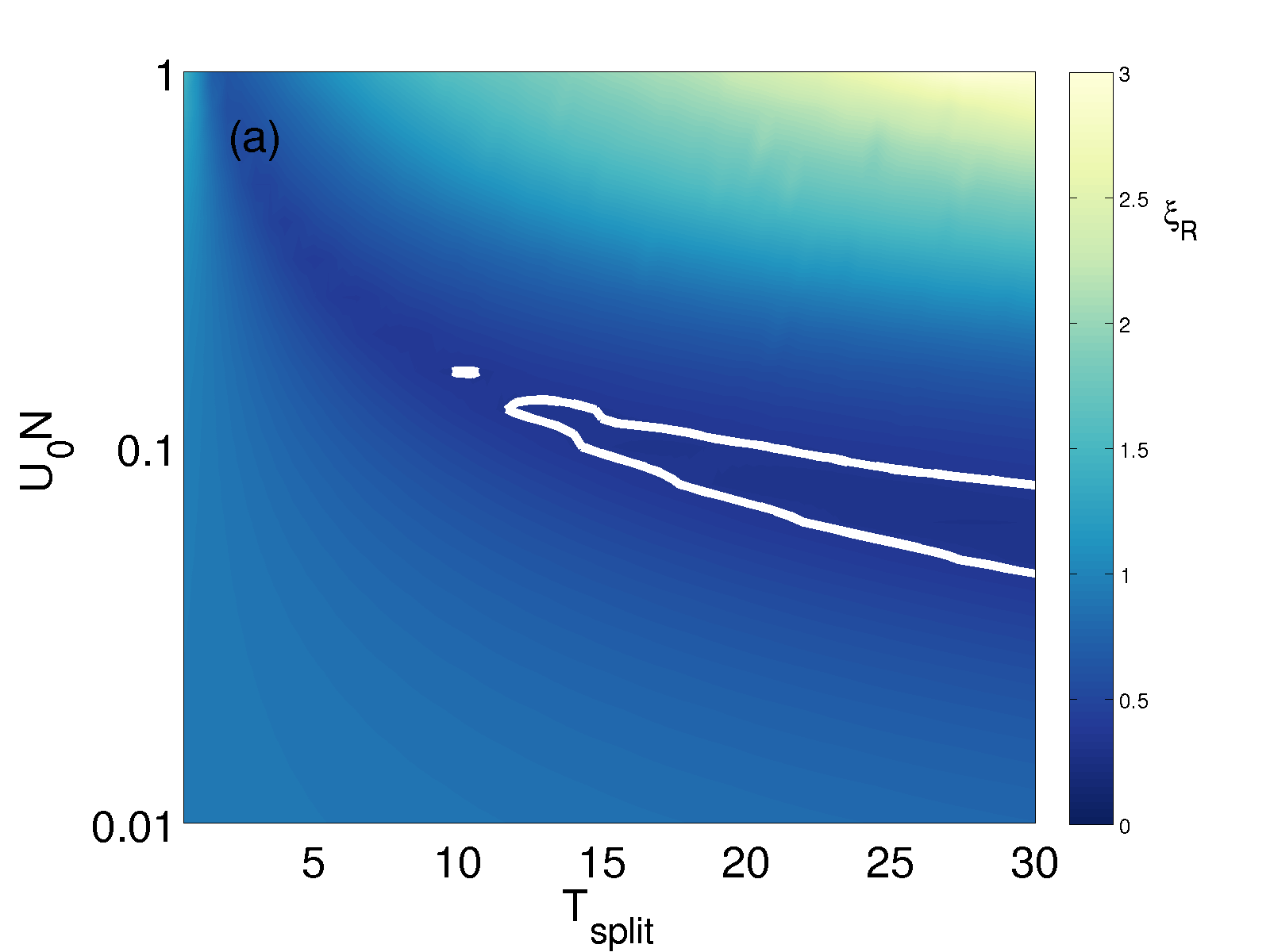}& \includegraphics[width=0.49\columnwidth]{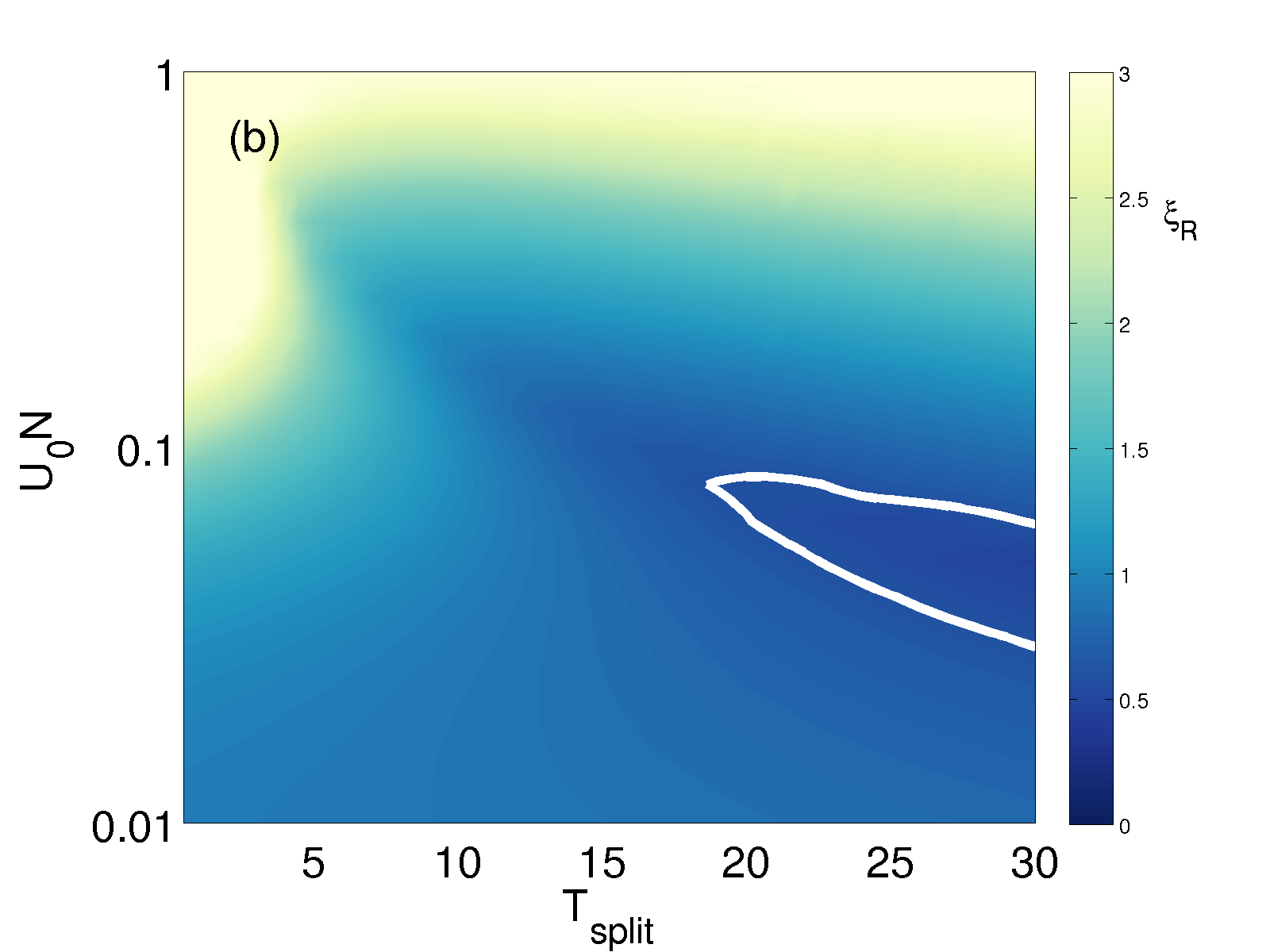}\\
  (a)&(b)
  \end{tabular}
\caption{(a,b) Phase sensitivity for a sequence including splitting and phase accumulation, versus interaction strength $U_0 N$ and $T_{\rm split}$. Parameters are $N=100$, (a) $T_{\rm phase}=1$, (b) $T_{\rm phase}=20$, and $T_{\rm pulse}=2$. For a given $T_{\rm split}$ we take the best number squeezing achieved by OCT. Number squeezing is higher for larger $T_{\rm split}$ and $U_0$ values. We also optimize for a tunnel pulse which rotates the number squeezed state.  The minimum phase sensitivity decreases very slowly with $T_{\rm split}$. The contour lines are at (a) $\xi_R=0.35$ and (b) $\xi_R=0.6$.
 \label{fig:kappaturnoff}}
\end{figure}
The best possible phase sensitivity for various $T_{\rm split}$ and $U_0 N$ values is shown in figure~\ref{fig:kappaturnoff} for (a) $T_{\rm phase}=1$ and (b) $T_{\rm phase}=20$.  For both cases sensitivity distinctly sub-shot noise sensitivity can be achieved. To achieve the squeezing needed to boost interferometer performance, a finite $U_0$ is needed, and for a given $T_{\rm split}$ there exists an optimal value of $U_0$. This value decreases for longer splitting times. 

In order to analyze the dependence on $N$, we consider a realistic 3D cigar-shaped trap with transverse trapping frequency $\omega_{\perp}=2\pi\times 2$ kHz as typically realized in atom chip interference experiments.  The effective 1D interaction strength in the splitting direction $U_0$ is then approximately proportional to $C_{1D}^{2/5}$. A more rigorous estimate, that is used in the calculations, is given in Ref.~\cite{grond.pra:09b}. For $N=100$, $U_0 N\sim1(0.1)$ corresponds to an aspect ratio of $C_{1D}\sim1/100(1/1000)$. These values change for $N=1000$ to $C_{1D}\sim1/1000(1/10000)$. For the pan cake-shaped trap we have $C_{2D}\sim \sqrt{C_{1D}}$, and thus $U_0 N=0.01$ is within reach for $C_{2D}\sim 1/1000$ and $N=100$, $C_{2D}\sim 1/10000$ and $N=1000$. 

\begin{figure}
  \begin{tabular}{l l}
  \includegraphics[width=0.67\columnwidth]{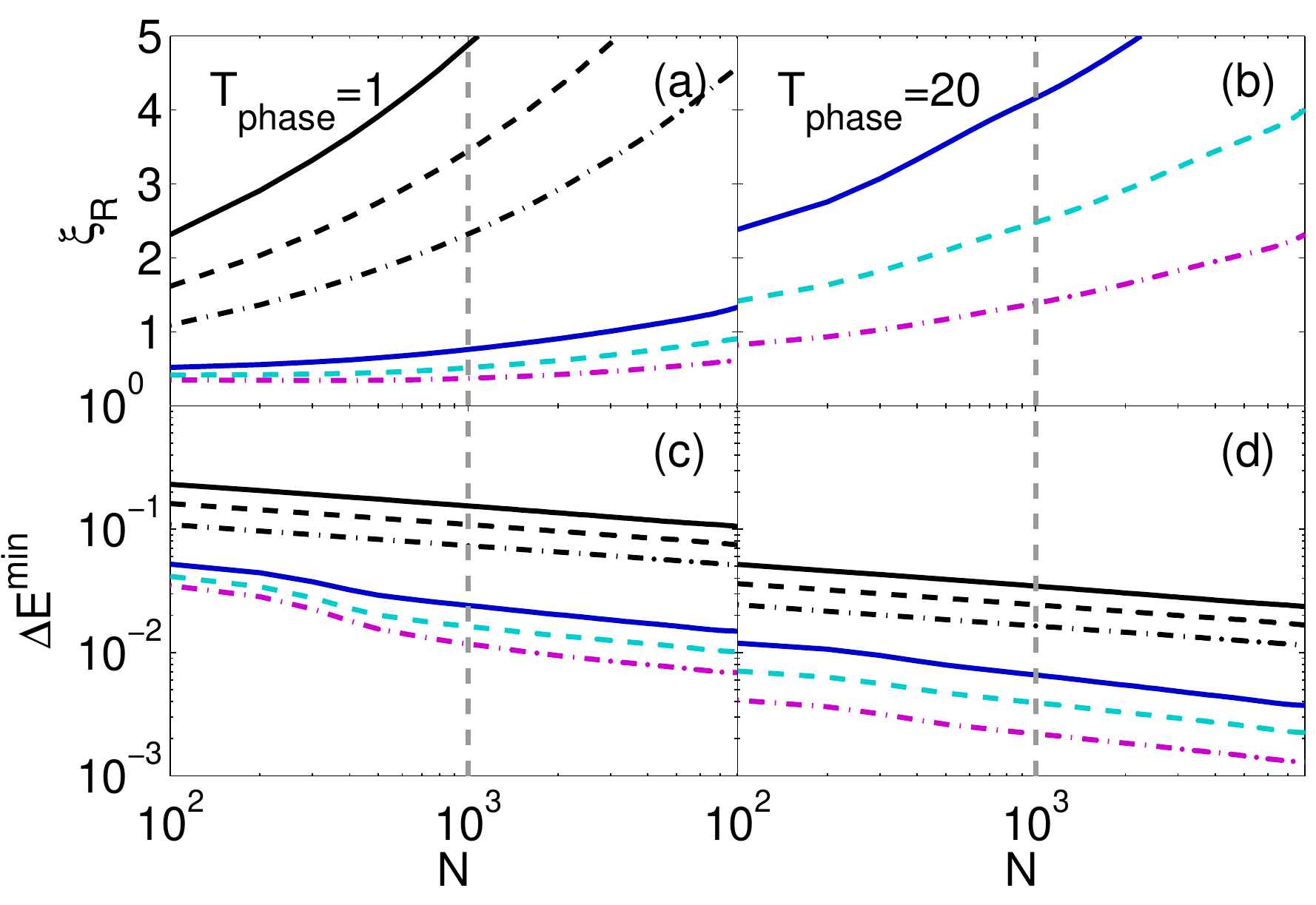}&  \includegraphics[width=0.33\columnwidth]{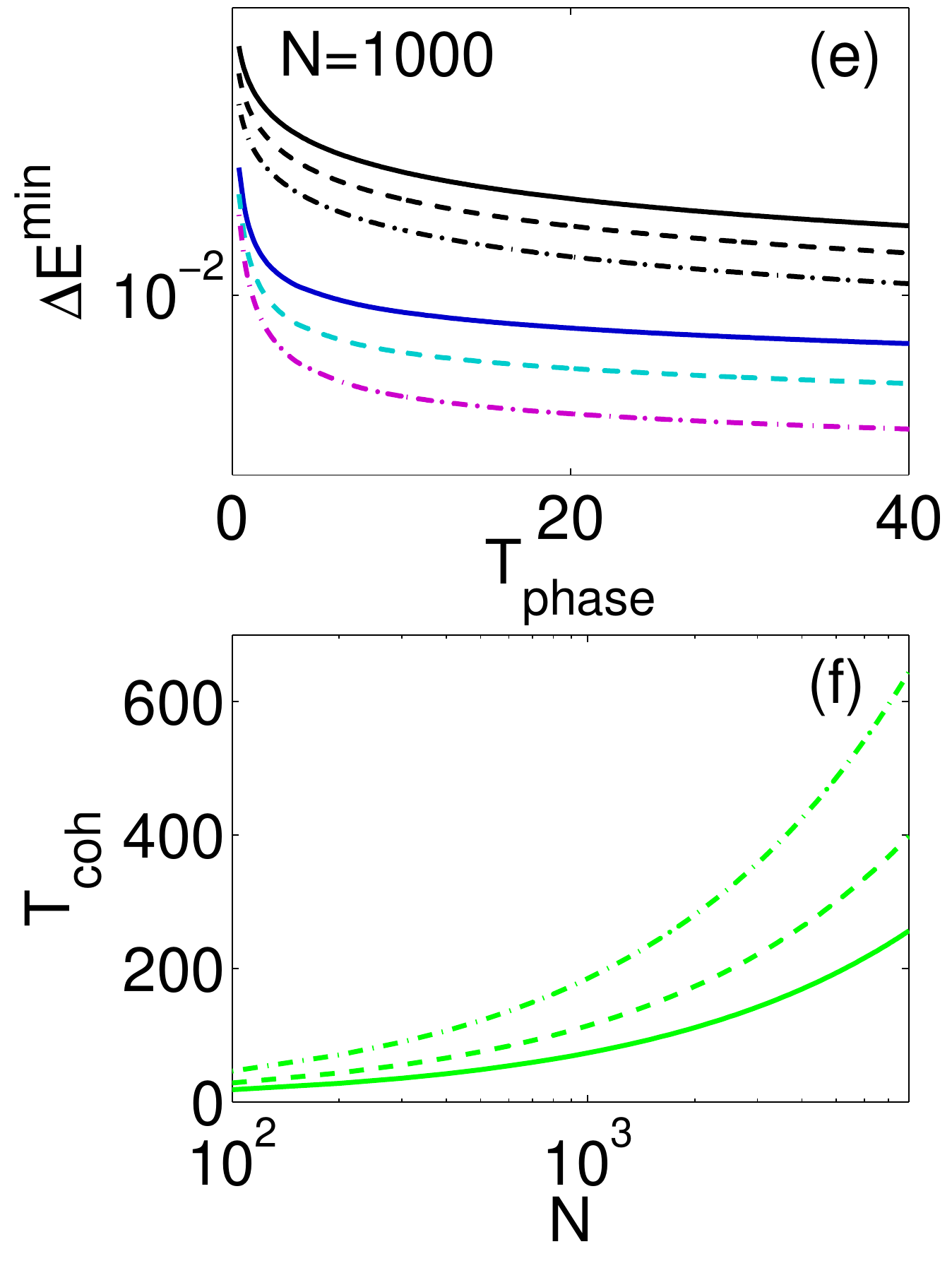}
  \end{tabular}
\caption{Optimization for a realistic cigar shaped trapping potential with transverse frequency $\omega_{\perp}=2\pi\cdot 2$ kHz ($\omega_{\perp}\approx 17 $ in scaled units) and aspect ratio $C_{1D}=1/100$ (solid lines), $C_{1D}=1/300$ (dashed lines), and $C_{1D}=1/1000$ (dashed-dotted lines). (a), (b) $\xi_R$ versus $N$ for (a) $T_{\rm phase}=1$ and (b) $T_{\rm phase}=20$.  (c), (d) Minimal detectable potential $\Delta E^{\rm min}$ versus $N$ for (c) $T_{\rm phase}=1$ and (d) $T_{\rm phase}=20$. (e) $\Delta E^{\rm min}$ versus time for $N=1000$ (the vertical dashed lines in (a)-(d) denote cuts at $N=1000$). The black lines show results for the case when only the initial number squeezing is optimized. The coloured lines show results with refocusing, i.e., where both initial number squeezing and tunnel pulse are optimized.   (f) Time $T_{\rm coh}$ after which $\xi_R$ is expected to grow rapidly due to loss of coherence, if refocusing is not applied.
\label{fig:Vmin}}
\end{figure}

Let us first consider the case without refocusing. We find approximately $\xi_R \sim 2\sqrt{2U_0 N T_{\rm phase}}$, and, considering the dependence of $U_0$ on the trapping potential, we obtain 
\be\label{eq:estxiN}
\xi_R \sim a_s^{1/5} \omega_{\perp}^{3/5} C_{1D}^{1/5}N^{1/5}T_{\rm phase}^{1/2}\;.
\ee
In order to reach sub-shot noise phase sensitivity we need the confinement ratio $C_{1D}$, atom number $N$, and $T_{\rm phase}$ all to be small.  In figure~\ref{fig:Vmin} (a) $\xi_R$ is plotted for $C_{1D}=1/100,1/300$ and $1/1000$ (solid, dashed, and dashed-dotted line, respectively). 

As we have seen in section~\ref{subsec:results}, equation~\eqref{eq:estxiN} is valid only if the coherence is well preserved. We estimate breakdown of this approximation when $\alpha\approx1-\frac{\xi_N^2}{2N}\approx 0.6$. From this we can obtain the time after which $\xi_R$ is expected to grow rapidly because the coherence factor tends to zero. It is given as $T_{\rm coh}=\frac{N^{3/5}}{2\cdot15^{2/5} a_s^{2/5} \omega_{\perp}^{6/5}C_{1D}^{6/15}}$, and shown in figure~\ref{fig:Vmin} (f) for different $C_{1D}$. 

We now turn to refocusing.  In the 1D elongated trapping geometry we find a more moderate increase of $\xi_R$ with N compared to the case without refocusing, see figures~\ref{fig:Vmin} (a) and (b).  This illustrates that the refocusing works better for large N, as long as $U_0 N$ is constant, see also figure~\ref{fig:figxi}.

The accumulated phase for a potential $\Delta E$ is given as $\theta=\Delta E T_{\rm phase}$, and the smallest detectable potential difference in time $T_{\rm phase}$ becomes $\Delta E^{\rm min}={\xi_R}/({\sqrt{N} T_{\rm phase}})$. Without refocusing, we find an improvement with N and $T_{\rm phase}$, $\Delta E^{\rm min}\sim N^ {-1/3}T_{\rm phase}^{-1/2}$, figure~\ref{fig:Vmin} (c), (d) and (e). Similar scalings also holds for the case with refocusing. We expect that for 2D traps, $\Delta E^{\rm min}$ somewhat below $10^{-3}$ is within reach. 

This confirms, in agreement with the scaling analysis of section~\ref{sec:opt.simple}, that a stronger trap anisotropy appreciably helps to reduce phase diffusion and, in turn, to improve the phase sensitivity of the interferometer. Possible limitations of strongly anisotropic systems are discussed in Sec.~\ref{sec:temp}.
The atom number $N$ helps to improve absolute sensitivity, but makes it more difficult to demonstrate measurements with a sensitivity below shot noise. 

\section{Interferometer performance within MCTDHB\label{sec:MCTDHB}}

Until now we have used a generic two-mode model to describe the interferometer, which captures the basic processes and physics, but ignores many details of the condensate dynamics in realistic microtraps. More specifically, the modeling of the splitting process and of the rotation pulses requires in many cases a more complete dynamical description in terms of the \emph{multi-configurational time dependent Hartree for Bosons} method (MCTDHB) \cite{alon:08}. In this section we discuss first the MCTDHB details relevant for our analysis, and its relation to the generic model. The main part will be concerned with the simulation and optimization of tunnel pulses for achieving tilted squeezed states, as discussed in section~\ref{subsubsec:spec} in context of refocusing. An exhaustive discussion of MCTDHB, as well as optimal condensate splitting can be found elsewhere \cite{alon:08,grond.pra:09b}.

In the two mode Hamiltonian of equation~\eqref{eq:hamtwomode} we did not explicitly consider the shape of the two {\textit{orbitals} $\phi_L$ and $\phi_R$, but lumped them into the effective parameters $\Omega$ and $\kappa$. The dynamics is then completely governed by the wavefunction accounting for the atom number dynamics. Within MCTDHB, both the orbitals and the number distribution are determined self-consistently from a set of coupled equations, which are obtained from a variational principle. This leads us to a more complete description, accounting for the full condensates' motion in the trap. The state of the system is then given by a superposition of symmetrized states (permanents), which comprise the time dependent orbitals. Instead of left and right orbitals, such as used in the two-mode model, we employ for the symmetric confinement potential of our present concern orbitals with \emph{gerade} and \emph{ungerade} symmetry.  The time dependent orbitals then obey non-linear equations, which depend on the one- and two-particle reduced densities \cite{sakmann:08} describing the mean value and variances of the number distributions \cite{streltsov:06}. 

The atom number part of the wavefunction obeys a Schr\"odinger equation with the Hamiltonian 
\begin{equation}\label{eq:TMMCham}
\mathcal{H}=\Omega\hat{J}_x+\frac{1}{2}\sum_{k,q,l,m}\hat a_k^{\dagger}\hat a_q^{\dagger}\hat a_l\hat a_m W_{kqlm}\,,
\end{equation}
where the indices are either $g$ (\emph{gerade}) or $e$ (\emph{ungerade} or excited).  We observe that, in contrast to the two-mode Hamiltonian of equation~\eqref{eq:hamtwomode}, the atom-atom interaction elements
\begin{equation}\label{eq:mvs}
W_{kqlm}=U_0\int dx \phi_k^*(x,t)\phi_q^*(x,t)\phi_l(x,t)\phi_m(x,t)\,,
\end{equation}
as well as the tunnel coupling $\Omega=\int dx \phi_e^*(x)\hat{h}\phi_e(x)-\int dx \phi_g^*(x)\hat{h}\phi_g(x)$ are governed by the orbitals. The only input parameter of the MCTDHB approach is the trapping potential $V_{\lambda}(x)$, which enters in the single-particle Hamiltonian $\hat h(x)=-(\nabla^2/2)+V_{\lambda}(x)$. We note that $\Omega$ obtained within MCTDHB cannot be directly interpreted as tunnel rate, but has to be renormalized if the two-body matrix elements differ from each other \cite{javanainen:99,ananikian:06}. Thus, there is in general no direct correspondence between the two-mode model and the MCTDHB approach. MCTDHB, which relies on time-dependent orbitals, captures a large class of excitations not included in a two-mode model. If the calculations converge when using more modes, MCTDHB reproduces the exact quantum dynamics, as discussed in \cite{sakmann:09,streltsov:09,grondscrinzi:10}.

In our MCTDHB calculations we consider a cigar-shaped magnetic confinement potential prototypical for atom chips \cite{folman:02}. Splitting is assumed to be along a transverse direction and is accomplished using rf dressing \cite{lesanovsky:06,lesanovsky:06b,hofferberth:06,hofferberth:07b}. For illustration purpose, the trapping potential in splitting direction can to very high accuracy be described by a quartic potential of the form:
\be
V(x,t) = a(t) x^2 + b(t) x^4\;,
\ee
where for most cases $b(t)$ varies very slowly and can be assumed as constant. This potential grasps the essential features of the initial and final potential and the time evolution $a(t)$ describes how the potential is split and the barrier is ramped up. $a(t)$ large and positive characterize the initial single  well, $a(t)$ large and negative the split double well, the constant $b(t)$ the confinement during the splitting.  In the calculations we use the exact form of the potential used in atom chip double well experiments \cite{lesanovsky:06}.  To describe the transformation of the potential we introduce a control parameter $\lambda(t)$ connected to the amplitude (phase) of the RF field.  Thereby, values of $-2/3<\lambda<0$ translate to a  single well, and $\lambda\sim 1$ to a double well.

Within MCTDHB the pseudospin operator $\hat J_x$ has to be rewritten in terms of the \emph{gerade} and \emph{ungerade} orbitals. In the new basis it measures the atom number difference with respect to the two states, $\hat{J}_x=\frac 1 2 (\hat a_g^{\dagger}\hat a_g -\hat a_e^{\dagger}\hat a_e)$, as discussed in more detail in the appendix.  It is important to note that the \emph{gerade} and \emph{ungerade} orbitals are \emph{natural} orbitals, i.e., they diagonalize the one-body reduced density of the system \cite{sakmann:08}. Therefore, if both of them are macroscopically populated one obtains a \emph{fragmented} condensate \cite{leggett:01}. We can thus interpret the coherence factor $\alpha$ as the degree of fragmentation. The system has maximal coherence, if its state is \emph{not} fragmented but forms a single condensate. Coherence is lost if the condensate fragments into two independent condensates. In between, we have a finite, but reduced coherence. Similarly, we can interpret $\Delta J_x$ as the number uncertainty between the fragmented parts. 

Optimal control theory (OCT) \cite{peirce:88,hohenester.pra:07}, is a very powerful tool to find a path which optimizes for a certain control target. In our earlier work \cite{grond.pra:09,grond.pra:09b} we implemented and optimized condensate splitting within MCTDHB \cite{peirce:88,hohenester.pra:07}. We found that, although the generic two-mode model describes qualitatively the splitting dynamics, the more complete MCTDHB description is needed for a realistic modeling. This is because one needs to control condensate oscillations during splitting, and to ensure a proper decoupling of the condensates at the end of the control sequence. In this section we employ OCT to find the appropriate paths in varying the trapping potential to achieve the desired \emph{tilting} on the Bloch sphere in short time $T_{\rm pulse}$ and prevent excitation of condensate oscillations.

\subsection{Parameter correspondence between the models \label{subsec:Corr}}

\begin{figure}
\center\includegraphics[width=0.75\columnwidth]{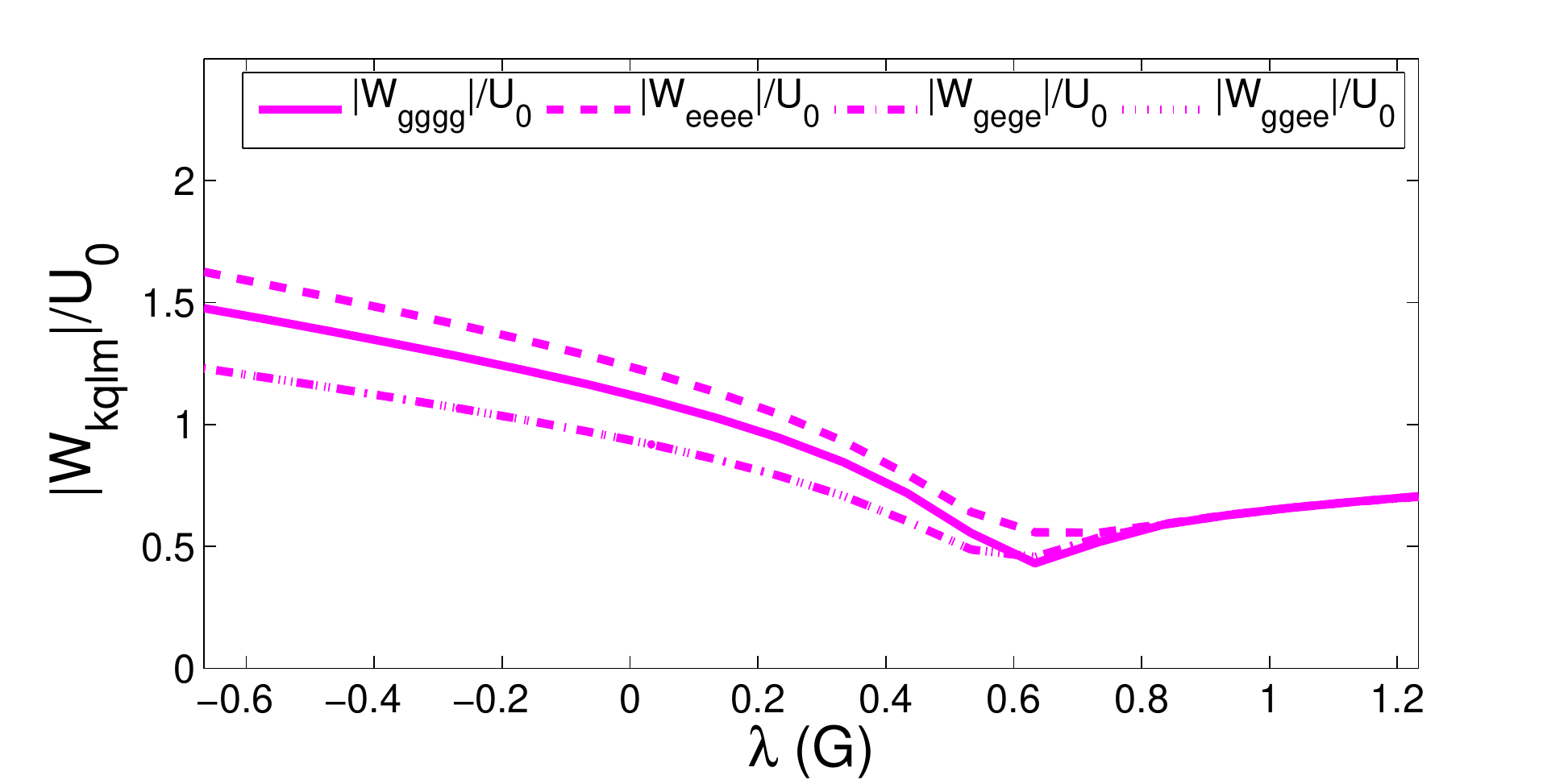}
\caption{ Two-body matrix elements for the ground states of a magnetic trap \cite{lesanovsky:06} versus splitting distance for $N=100$ and $U_0 N=1$. The splitting is parametrized by the parameter $\lambda$. $W_{gege}$ and $W_{ggee}$ coincide for the ground states, however not in general, see the examples of figures~\ref{fig:demoOCT1} and \ref{fig:demoOCT2}. \label{fig:mv}}
\end{figure}

For the calculation of time-dependent condensate dynamics including oscillations, a self-consistent approach like MCTDHB is mandatory. However, we expect the generic model (equation~\eqref{eq:hamtwomode}) to properly describe the phase accumulation stage ($\Omega=0$), provided the condensates are at rest. The two-body overlap integrals of the orbitals from MCHB (time-independent version of MCTDHB \cite{streltsov:06}), given in equations~\eqref{eq:mvs}, are then constant and coincide. This is because $\phi_g$ and $\phi_e$ have degenerate moduli for split condensates. When comparing equations~\eqref{eq:hamtwomode} and \eqref{eq:TMMCham}, we find the value of $\kappa$ to be used in the generic model. Similarly, in the context of optimized splitting protocols in our earlier work \cite{grond.pra:09,grond.pra:09b}, we have found that both optimizing $\Omega(t)$ in the generic model and optimizing $\lambda(t)$ within MCTDHB yield the same amount of number squeezing for a given time interval. 

 In figure~\ref{fig:mv} we show how the two-body overlap integrals from MCHB vary with the control parameter that determines the shape of the confinement potential. During the transition from a single-well to a double well, they drop by roughly a factor of two. This is because in the final state the \emph{gerade} and \emph{ungerade} orbitals are delocalized over both wells. After reaching a minimum around $\lambda\sim0.7$, the overlap integrals start to increase again slightly. In context of splitting we found that it is reasonable to assume $\kappa=U_0/2$ throughout the splitting process \cite{grond.pra:09b}, i.e., to take the value in the most relevant regime during condensate breakup ($\lambda\sim0.7-0.8$). Atom interferometry has to be performed with split condensates. This requires a splitting distance of some micrometers, corresponding to $\lambda \gtrsim 1$. With the corresponding value of $\kappa\approx 0.65 U_0$, phase diffusion in the phase accumulation stage can be well described using the generic model. This has in particular the advantage that we can translate our findings for the optimal states for atom interferometry from section \ref{subsubsec:spec} to MCTDHB calculations of tunnel pulses.

\subsection{Pulse optimization}

A key ingredient in interferometer performance is the preparation of an optimized initial state, as discussed in section~\ref{subsubsec:spec}.  This can be achieved by a 'tunnel pulses' facilitating the \emph{tilting} of the initial number squeezed state. This operation has been analyzed by Pezz\'e \emph{et al.} in the context of a cold atom beam splitter.  They used the generic model \cite{pezze2:06} and studied to which extent the creation of phase squeezed states from number-squeezed states is spoiled by atom-atom interactions. The real space dynamics has been neglected. 

In real space, the confining potential has to be modified to bring the condensates together for the tunnel pulse which accomplishes the desired 'tilt' on the Bloch sphere.   As has been discussed in section~\ref{subsec:opt_trap}, the duration of the tunnel pulse ($T_{\rm pulse}$) is very critical for the interferometer performance, the pulse should be as short as possible. This has to be done without significantly disturbing the many-body wave function.  To design appropriate control schemes with the shortest possible time duration we will have to take the real space dynamics of the BEC into account using MCTDHB. 

To find a control sequence for the preparation of the optimal initial states at $T_{\rm phase}=0$ for a given interferometer sequence within MCTDHB, we first start, following section~\ref{subsec:Corr}, with the optimized initial number squeezing and the tilt angle required for rephasing at a given $T_{\rm phase}$ as calculated in section~\ref{subsubsec:spec}  within the generic model.  We then choose the top of a Gaussian shaped initial guess 
\begin{equation}\label{eq:pulseguess}
\lambda(t)=\lambda_0+(\lambda_{\rm end}-\lambda_0)\cdot t/T_{\rm pulse}-A\Bigl[e^{-\frac{(t-T_{\rm pulse}/2)^2}{2 B^2}}-e^{-\frac{(T_{\rm pulse}/2)^2}{2 B^2}}\Bigr]\,.
\end{equation}
We start with a state of the double well potential $V_{\lambda_0}$ with the required initial number squeezing.  Then, as $\lambda$ decreases, the barrier is ramped down, the condensates approach each other. The desired $\Delta J_y^d$ to be reached at $T_{\rm pulse}$ is fixed by the required tilt angle. It can be tuned by the parameters $A$ and $B$, corresponding to the depth and the width of the control parameter deformation, respectively. Finally, a double well $V_{\lambda_{end}}$ is re-established, which completely suppresses tunneling of the two final condensates, at least if they are in the ground state.

Results of our MCTDHB calculations are shown in figures~\ref{fig:demoOCT1} and \ref{fig:demoOCT2} for interactions $U_0 N=0.1$ and $U_0 N=1$, respectively.  The pulse achieves the desired $\xi_R$ [dashed lines in (d)], as we expected from the generic model (dashed-dotted lines). However, it not only affects the atom number distribution, but also leads to an oscillation of the condensates in the microtrap. This can be seen in the density, which is depicted in figures~\ref{fig:demoOCT1}(b) and \ref{fig:demoOCT2}(b). Condensate oscillations during the phase accumulation and release stage are expected to substantially degrade the interferometer performance, and may even lead to unwanted condensate excitations \cite{grondscrinzi:10}. These oscillations can be avoided by using more refined \emph{tilting} pulses, which can be obtained within the framework of OCT, where we now optimize for phase squeezing and a desired $\Delta J_y^d$ at the final time of the control interval $T_{\rm pulse}$, corresponding to $T_{\rm phase}=0$. Some details of our approach are given in the appendix.

\begin{figure}
\center\includegraphics[width=0.9\columnwidth]{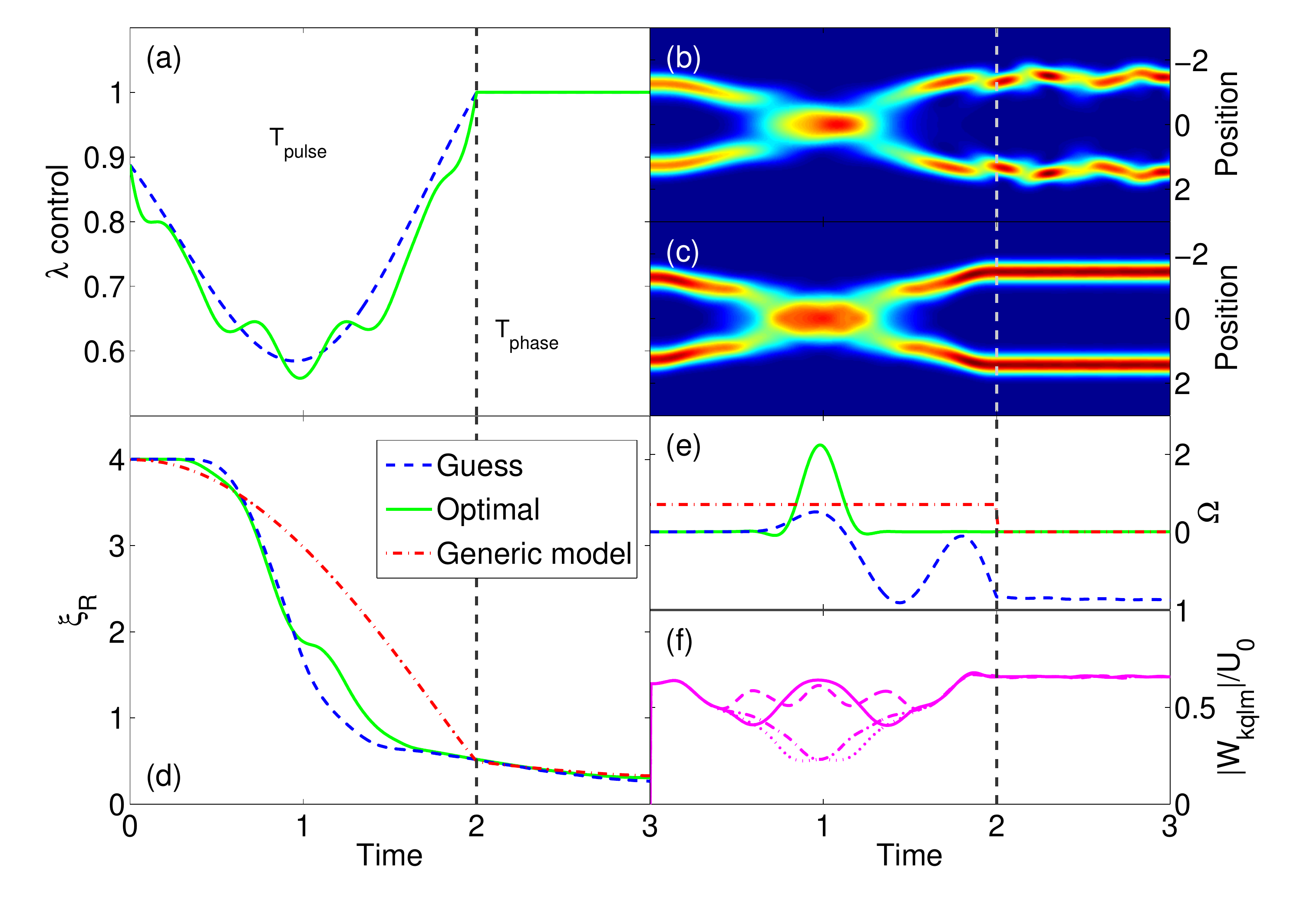}
\caption{(a) Typical OCT-control (green solid line) for weak interactions $U_0 N=0.1$, atom number $N=100$, $T_{\rm pulse}=2$ and $T_{\rm phase}=1$, compared to a simpler control (blue dashed line). The dashed vertical line separates the control sequence from the phase accumulation time thereafter. The corresponding condensates density are shown (b) for the initial guess and (c) for the OCT-solution, and we compare (c) useful squeezing $\xi_R$ and (d) tunnel coupling $\Omega$. Additionally we compare to results from the generic two-mode model, where a square-$\Omega$ pulse is used (red dashed-dotted lines). (f) The two-body matrix elements of the OCT-solution are shown, similar as in figure~\ref{fig:mv}. \label{fig:demoOCT1}}
\end{figure}

For weak interactions a short tunnel pulse $T_{\rm pulse}=2$ can be easily found which properly tilts the atom number distribution and brings the condensates to a stationary state at the end of the process.  A typical control sequence for $U_0 N=0.1$ and optimized for a short $T_{\rm phase}=1$ is shown in figure~\ref{fig:demoOCT1}(a) [green solid line], the corresponding density given in (c). In panel (d) we depict the phase sensitivity $\xi_R$ (solid line), which compares in the phase accumulation stage very well with the desired behavior given by the generic model (dashed-dotted line).

The gain of our OCT solution with respect to an `adiabatic' control of Gaussian type [equation~\eqref{eq:pulseguess}], where $\lambda$ is modified sufficiently slowly in order to suppress condensate oscillations, depends on the chosen value of $T_{\rm phase}$. Smaller $T_{\rm phase}$ values require larger tunnel pulses [see figure~\ref{fig:figxi}(c)]. In our example, the OCT pulses can be at least one order of magnitude faster, which means an improvement of up to $30 \%$ in $\xi_R$ [compare also with figure~\ref{fig:Tpulse}]. In figure~\ref{fig:demoOCT1}, panels (e) and (f), we report the tunnel coupling and two-body matrix elements.  For the initial guess, which has a wildly oscillating density, also the tunnel coupling oscillates strongly and takes on a finite value after the control sequence.  We interpret this as a signature of condensate excitations, which go beyond the two-mode MCTDHB model. In contrast, a smooth tunnel pulse and stationary final condensates are achieved for the optimal control. MCTDHB simulations with a higher number of modes indicate that the two-mode approach provides a very accurate level of description \cite{grondscrinzi:10}. The two-body matrix elements of the optimized solutions show a complex behaviour and deviate from each other quite appreciably, which demonstrates that the dynamics depends critically on the orbitals [see equation~\eqref{eq:mvs}]. 

\begin{figure}
\center\includegraphics[width=0.9\columnwidth]{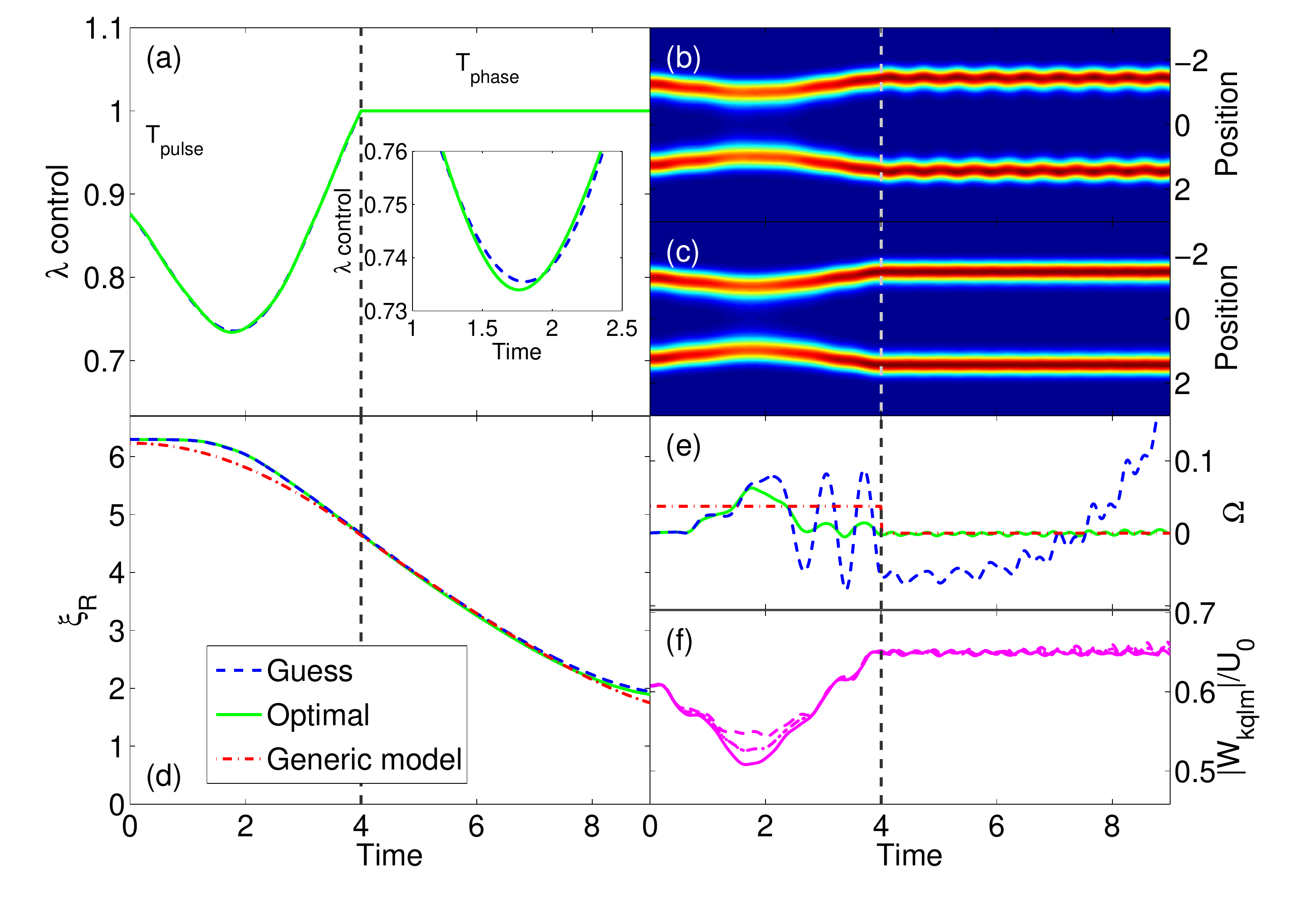}
\caption{Same as figure~\ref{fig:demoOCT1}, but for $U_0 N=1$, $T_{\rm pulse}=4$ and $T_{\rm phase}=5$. (a) The inset magnifies the controls. (b)-(f) describe the same quantities as in figure~\ref{fig:demoOCT1}. \label{fig:demoOCT2}}
\end{figure}

For stronger interactions, we find that it becomes more difficult to decouple the condensate at the end of the \emph{tilting} pulse, as shown in figure~\ref{fig:demoOCT2} for $U_0 N=1$ and $T_{\rm phase}=5$. Although we easily achieve trapping of the orbitals in a stationary state, also for shorter $T_{\rm pulse}$ or $T_{\rm phase}$, we find that number fluctuations are not constant after the control sequence (results are not shown), which is a signature of additional unwanted condensate excitations \cite{grondscrinzi:10}. These excitations are most pronounced when the coherence factor tends to one, and the \emph{ungerade} orbital becomes depopulated. In particular, we could not find pulses shorter than $T_{\rm pulse}=4$ which lead to a final decoupling of the condensates. 
The same holds for the initial guess (blue dashed lines in figure~\ref{fig:demoOCT2}). Although the condensate oscillations after the pulse sequence are moderate, they do not uncouple, as indicated by the growing tunnel coupling at later times. In contrast, the OCT control [green solid line in panel (a), magnified in the inset] achieves decoupling and a complete suppression of any tunnelling. Also the two-body matrix elements are stationary to a good degree. 

In conclusion, we find that optimized \emph{tilting} pulses are crucial for avoiding condensate oscillations in the phase accumulation stage, and for reducing unwanted condensate oscillations. Adiabatic pulses might be orders of magnitude longer, and thus appreciably reduce the interferometer performance. 

\section{Influence of temperature on the coherence of interferometer measurements \label{sec:temp}}

Up to now we considered the atom cloud (BEC) at zero temperature.  In realistic experiments the quantum system will be at some finite temperature, and especially for the favorable low dimensional geometries temperature effects and decoherence due to thermal or even quantum fluctuations might become important \cite{Petrov2004,Cazalilla2004,Imambekov2007,Bistritzer2007}. We now turn to look at the effects of temperature on the decoherence of a split BEC interferometer. 

\subsection{One-dimensional systems}

In one dimensional quantum systems ($k_B T$ and $\mu \; < \; \hbar \omega_\perp$) fundamental quantum fluctuations prevent the establishment of phase coherence in an infinitely long system even at zero temperature \cite{Cazalilla2004}.  The coherence between two points in the longitudinal direction $z_1$ and $z_2$ decays as $ \left|z_1 - z_2 \right|^{-1/2{\cal K}} $, where ${\cal K} = \pi \hbar \sqrt{{n_{1D}}/{(g_{1D} m)}}= \pi n_{1D} \zeta_h$ is the Luttinger parameter.  Thereby $\zeta_h = \hbar / \sqrt{m n_{1D} g_{1D}}$ is the healing length, $n_{1D}$ is the 1D density, and $g_{1D}=2 \hbar \omega_\perp a_s$ is the 1D coupling strength in a system with transversal confinement $\omega_\perp$ and scattering length $a_s$. For a weakly interacting 1D system ${\cal K} \gg 1$, and the length scale where quantum fluctuations start to destroy phase coherence is
\be
l_\Phi ^{quant} \approx \zeta_h e^{2 {\cal K}}\;. 
\ee
For weakly interacting one dimensional Bose gases $\zeta_h$ is in the order of 0.1 to 1 $\mu$m. With $\mathcal K > 10$ one can safely neglect quantum fluctuations (see also  \cite{Bistritzer2007}).


These thermal phase fluctuations are also present in a very elongated 3D Bose gas. In such a finite 1D system one can achieve phase coherence (i.e., $l_\Phi ^{therm}$ becomes larger than the longitudinal extension of the atomic cloud) if the temperature is below $T_\Phi = T_{1D} \frac{\hbar \omega_z}{\mu} \approx n_{1D} \zeta_h  \hbar \omega_z$.  To achieve a practically homogeneous phase along the BEC $T < T_\Phi/10$ is desirable \cite{Petrov2004}.

For interferometry only the relative phase between the two interfering systems and its evolution are important. After a coherent splitting process, even a phase fluctuating condensate is split into two copies with a uniform relative phase.  For interferometer measurements decoherence of this definite relative phase is adverse. 

The loss of coherence in 1D systems due to thermal excitations was considered theoretically by Burkov, Lukin and Demler \cite{burkov:07}, and by Mazets and Schmiedmayer \cite{Mazets2009}, and probed in an experiment by Hofferberth \emph{et al.} \cite{hofferberth:07} and Jo \emph{et al.} \cite{jo:07}.  The coherence in the system left is characterized by the coherence factor 
\be
\Psi(t) = \left\langle \frac{1}{L} \left| \int_0 ^L dz \; e^{i \theta(z,t)}\right| \right\rangle\;,
\ee
where $\theta(z,t)$ is the relative phase between the two condensates. The angular brackets denote an ensemble average.  The key feature in both calculations is that for 1D systems the coherence factor $\Psi(t)$ decays non-exponentially :
\be
\Psi(t) \propto \exp\left[{-(t/t_0)^{2/3}}\right]\;.
\ee
Burkov, Lukin and Demler \cite{burkov:07} give for the characteristic time scale 
\be
t_0 = 2.61 \pi \frac{\hbar \mu}{T^2} \; {\cal K}\;,
\ee
whereas Mazets and Schmiedmayer  \cite{Mazets2009} find: 
\be
t_0 = 3.2 \frac{\hbar \mu }{T^2} \left(\frac{{\cal K}}{\pi}\right)^2 = 3.2 \frac{\hbar^3 n_{1D} ^2}{m T^2}\;.
\ee
Even though the two show a different scaling with the Luttinger parameter $\cal K$, both are consistent with Hofferberth \emph{et al.} \cite{hofferberth:07} within the experimental error bars in the probed range of ${\cal K} \sim 30$.  

There are two strategies to get long coherence times: 
\begin{itemize}
	\item The characteristic timescale for decoherence scales like $T^{-2}$ indicating that very long coherence times can be reached for low temperatures. 
	\item The time scale of decoherence scales with the 1D-density as  $t_0 \propto n_{1D} ^2$ in \cite{Mazets2009}, and $t_0 \propto n_{1D} ^{3/2}$ in \cite{burkov:07}. Increasing the 1D density will enhance the coherence time available for the measurement.
\end{itemize}
Putting all together coherence times $t_0$ in the order of $10^4 \hbar /\mu$ can be achieved for $T \approx 20$ nK and $n_{1D} \approx~ 60$ atoms/$\mu m$.

The above estimates have also to be taken with care, especially since the 1D calculations are only strictly valid for $T < \mu$, but we believe that they are a reasonable estimate also for elongated 3-d systems. 

The question is if the required low temperatures ($T \sim \mu/5$ in the above example) can be reached in 1D systems where two-body collisions are frozen out? In the weakly interacting regime, that is for small scattering length and relatively high density, this should be possible, because thermalization can be facilitated by virtual three body collisions \cite{Mazets2008,Mazets2009b,Mazets2010}.  In the experiments by Hofferberth \emph{et al.} $T \approx 30$ nK and $n_{1D} \approx~ 60$ atoms/$\mu m$ were achieved \cite{hofferberth:08}

Another point are dynamic excitations in the weakly confined direction. The interaction parameter $U_0$ in general varies when $\lambda$ is changed and this can lead to longitudinal excitations. We find numerically that this effect is very small for the trapping potentials considered. In addition one can in principle account for the effects of the change in interaction energy by controlling the longitudinal confinement. 

\subsection{Two-dimensional systems}

Quantum fluctuations are not important in 2D systems since a repulsive bosonic gas exhibits true condensate at $T=0$.

In a two dimensional Bose gas at finite temperature the phase fluctuations scale logarithmically with distance ($r\gg \lambda_T$) \cite{Petrov2004}: 
\be
 \left\langle \left[ \varphi(r)-\varphi(0) \right]\right\rangle_T \approx \frac{2 T}{T_{2D}} \ln \left(\frac{r}{\lambda_T}\right)\;,
\ee
where $T_{2D}=2 \pi \hbar^2 n_{2D}/m$ is the degeneracy temperature for a 2D system, $n_{2D}$ being the peak 2-d density.  For $T \gg \mu$ the length scale is $\lambda_T \approx \zeta_h$, where $\zeta_h = \hbar / \sqrt{m n_{2D} g_{2D}}$ is the healing length in the 2D system with the effective coupling $g_{2D}$. For weak interactions it is given by $g_{2D}=\frac{\sqrt{8 \pi} \hbar^2 a_s}{m \, l_\perp}$ with $l_\perp=\sqrt{\frac{\hbar}{m \, \omega_\perp}}$ \cite{Petrov2004}.  For $T \ll \mu$ one finds the length scale $\lambda_T \approx \frac{\hbar \, c_s}{T}$, which is equal to the wavelength of thermal phonons. The phase coherence length $l_\Phi$ is given by the distance where the mean square phase fluctuations become of the order of unity.
\be
l_\Phi \approx \lambda_T \exp \left(\frac{T_{2D}}{2 T}\right)\; .
\ee

The decoherence in 2D systems was considered by Burkov, Lukin and Demler  \cite{burkov:07}, where they find a power law decay of the coherence factor:
\be
\Psi(t) \propto t^{-T/8 T_{KT}}\;,
\ee   
for times larger then
\be
t_0 = \frac{3 \sqrt{3}}{4 \pi} \frac{\mu T_{KT}}{T^3}\;,
\ee
where $\mu = n_{2D} g_{2D}$ is the chemical potential of the 2D system. 
The temperature for the Kosterlitz-Thouless transition is given by 
\be
T_{KT} = \frac{\pi}{2} \frac{n_{2D} ^{sf}}{m} \leq \frac{1}{4}T_{2D}\;,
\ee
where $n_{2D} ^{sf} \leq n_{2D}$ is the super fluid density. Again at sufficiently low temperatures the decoherence due to thermal phase fluctuations is smaller than the phase diffusion due to the non-linear interactions during the phase accumulation stage.

\section{Summary and conclusions}

From our analysis of a double well interferometer for trapped atoms, it becomes evident that the main limiting factor to measurements with atom interferometers is the phase diffusion caused by the non linearity created by the atom-atom interactions. Consequently many of the recent experiments used interferometry to study the intriguing quantum many-body effects caused by interactions \cite{albiez:05,hofferberth:07,jo:07,hofferberth:08,esteve:08}.

Optimal control techniques can help improving interferometer performance significantly by designing optimized splitting ramps and rephasing pulses, but the overall performance of the interferometer is still limited by atom interactions, and not by the readout, except for experiments with very small atom numbers. In general low dimensional confinement of the trapped atomic cloud is better for interferometry. Nevertheless we found it difficult to get a performance for the minimal detectable shift $\Delta E_{\rm min} < 10^{-4} \mu$ even for an optimized setting with a 1D elongated trap. In addition we would like to point out that even though we did our analysis for a generic double well the same will hold for trapped atomic clocks \cite{treutlein:04}, where the signal comes from Ramsey interference of internal states.  For the internal state interferometers the difference of the interaction energies is the relevant quantity to compare. 

The most direct way to achieve a much improved performance is to decrease the atom atom interaction. The best is to cancel it completely by either putting the atoms in an optical lattice, where on each site the maximal occupancy is 1, or by tuning the scattering length $a_{s}=0$, which can in principle be achieved by employing Feshbach resonances \cite{chin:10}.  Drastic reduction of phase diffusion when bringing the scattering length close to zero was recently demonstrated in two experiments in Innsbruck \cite{gustavsson:08} and Firenze \cite{Fattori:08}. The big disadvantage thereby is that using Feshbach resonances requires specific atoms and specific atomic states.  These states need to be tunable, and are therefore \textit{not} the '\textit{clock}' states which are insensitive to external fields and disturbances.

In an ideal interferometer one would like to use clock states, create strong squeezing during the splitting process by exploiting the non-linearity in the time evolution due to atom-atom interactions, and then, after the splitting turn off the interactions (by setting the scattering length to $a_s = 0$).  All together might be difficult or even impossible to achieve.

For the interferometers considered here one can always reach low enough temperature to neglect decoherence due to thermal excitations even for 1D and 2D systems.

In addition to the interferometer scheme considered here, there exist other ideas of how atom interferometry could be improved. An interesting route will be to exploit in interferometry the correlations of the many boson states, and to establish a readout procedure which is immune to phase diffusion.  One approach is to use Bayesian phase estimation schemes for the analysis of the phase sensitivity \cite{pezze:07,pezze:09}. Other proposed schemes include monitoring the coherence and revival dynamics of the condensates \cite{dunningham:04}, the measurement of a phase gradient along the double well potential containing tunnel-coupled condensates by means of a contrast resonance \cite{jae:06}, or inhibition of phase diffusion by quantum noise \cite{khodorkovsky:08}. For those more advanced ways to read out the interference patterns, however,  the requirements for the temperature will become more stringent the better the new read-out schemes can compensate for the adverse effects of phase diffusion.

\section*{Acknowledgments}

We thank J. Chwede\'nczuk, F. Piazza, A. Smerzi, G. von Winckel, O. Alon, and T. Schumm for helpful discussions. This work has been supported in part by NAWI GASS, the FWF and the ESF Euroscores program: EuroQuaser project QuDeGPM.

\appendix
\section{\hspace{2.5 cm} Optimality system for pulse optimization}

In this appendix, we give details about the optimality system for pulse optimization. For this purpose, we first discuss the pseudo-spin operators of equation~\eqref{eq:pseudospin} within MCTDHB. Since the many-boson state depends on both the number distribution and the orbitals, the operators now depend on position and time. $\hat J_z$ measures the difference between atoms in the left and right well, and $\hat J_x$ the difference between atoms in \emph{gerade} and \emph{ungerade} orbitals. $\hat J_y$ is then determined such that the pseudospin operators fulfill the spin algebra. Then,
\begin{equation} \label{eq:pseudospin.mc}
\mbox{\hspace*{-2cm}}
\hat{J}_x=\frac 1 2 (\hat a_g^{\dagger}\hat a_g -\hat a_e^{\dagger}\hat a_e)\,,\quad\hat{J}_y=i (d\ah_g^{\dagger}\ah_e-d^*\ah_e^{\dagger}\ah_g)\,,\quad
\hat{J}_z=(d\ah_g^{\dagger}\ah_e+d^*\ah_e^{\dagger}\ah_g)\,,\quad
\end{equation}
where $d=\int_{-\infty}^{0}\phi_g^*(x)\phi_e(x)\,dx$ is the half sided overlap integral of the orbitals \cite{grond.pra:09b}. It measures the degree of orbital localization on the left-hand side. For our OCT optimization we define a cost functional for pulse optimization
\bea\label{eq:costMCTDHB}
\mbox{\hspace*{-2cm}}
J(\phi_g,\phi_e,\mathbf{C},\lambda) &=& \frac{\gamma_1}{2}\Bigl(\Delta J_y^d- \langle\mathbf{C}(T)|\hat{J}_y^2|\mathbf{C}(T)\rangle\Bigr)^2  \non
&&+ \frac{\gamma_2}{2}[2-|\la \phi_g(T)|\phi_g^d\ra|^2-|\la \phi_e(T)| \phi_e^d\ra|^2]
+ \frac{\gamma}{2}\int_0^T\bigl[\dot{\lambda}(t)\bigr]^2 dt\,,
\eea
where $\mathbf{C}$ is the atom number part of the wavefunction, and $\gamma_1$, $\gamma_2$, and $\gamma$ are weighting parameters. The first term quantifies the deviation of the actual $\Delta J_y$ from the desired one, the second term accounts for trapping the orbitals in the ground state, and the third term penalizes rapidly varying controls and renders the problem well posed. In our OCT approach, the cost function is minimized subject to the constraint that the many-boson wavefunction fulfills the proper MCTDHB equations. Details of this approach can be found in Ref.~\cite{grond.pra:09b}.

\section*{References}


\end{document}